\def\kB{k_{\text B}}

\def\Im{{\text{Im}}\,}

\def\kF{k_{\text{F}}}
\def\vF{v_{\text{F}}}
\def\NF{N_{\text{F}}}

\def\epsilonF{\epsilon_{\text F}}

\def\sgn{{\text{sgn\,}}}
\def\be{\begin{equation}}
\def\ee{\end{equation}}
\def\bea{\begin{eqnarray}}
\def\eea{\end{eqnarray}}
\def\bse{\begin{subequations}}
\def\ese{\end{subequations}}

\def\DT{D_{\text{T}}}
\def\perpgradT{({\hat{\bm k}}_{\perp}\!\!\cdot\!{\bm\nabla}T)}
\def\Ssym{S^{\text{sym}}}

\documentclass[prb,twocolumn,amsmath,amssymb,eqsecnum,preprintnumbers,floatfix]{revtex4}
\usepackage{graphicx}% Include figure files
\usepackage{dcolumn}% Align table columns on decimal point
\usepackage{bm}% bold math
\usepackage{bbm}
\usepackage{color}
\usepackage{enumitem}
\usepackage{soul}
\usepackage{bbold}
\begin{document}
%\preprint{J. Chem. Phys. B {\bf 125}, 7499 (2021)}

%\bibliographystyle{./prsty}

%\bibliographystyle{revtex4}

\title{A Fluctuation-Response Relation as a Probe of Long-Range Correlations in Non-Equilibrium Quantum and Classical Fluids}

\author{T.R. Kirkpatrick$^{1}$ and D. Belitz$^{2,3}$}

\affiliation{$^{1}$Institute for Physical Science and Technology, University of Maryland, College Park, MD 20742, USA\\
 $^{2}$Department of Physics and Institute for Fundamental Science, University of Oregon, Eugene, OR 97403, USA \\
 $^{3}$ Materials Science Institute, University of Oregon, Eugene, OR 97403, USA}

\date{\today}
\begin{abstract}
The absence of a simple fluctuation-dissipation theorem is a major obstacle for studying systems that are not in thermodynamic
equilibrium. We show that for a fluid in a non-equilibrium steady state characterized by a constant temperature gradient the
commutator correlation functions are still related to response functions; however, the relation is to the bilinear response of 
products of two observables, rather than to a single linear response function as is the case in equilibrium. This modified 
fluctuation-response relation holds for both quantum and classical systems. It is both motivated and informed by the long-range 
correlations that exist in such a steady state and allows for probing them via response experiments. This is of particular interest 
in quantum fluids, where the direct observation of fluctuations by light scattering would be difficult. In classical fluids it is known 
that the coupling of the temperature gradient to the diffusive shear velocity leads to correlations of various observables, in 
particular temperature fluctuations, that do not decay as a function of distance, but rather extend over the entire system. 
We investigate the nature of these correlations in a fermionic quantum fluid and show that the crucial coupling between the
temperature gradient and velocity fluctuations is the same as in the classical case. Accordingly, the nature of the long-ranged 
correlations in the hydrodynamic regime also is the same. However, as one enters the collisionless regime in the low-temperature 
limit the nature of the velocity fluctuations changes: they become ballistic rather than diffusive. As a result, correlations of the 
temperature and other observables are still singular in the long-wavelength limit, but the singularity is weaker than in the 
hydrodynamic regime. 
\end{abstract}

\maketitle

\section{Introduction}
\label{sec:I}

In a classical fluid, in equilibrium at a temperature $T_{\text{eq}}$ and far from any critical point, correlations are generically short-ranged, i.e.,
they decay exponentially on an atomic scale. In a coarse-grained description this corresponds to a $\delta$-function
correlation. Specifically, the correlations of the spatial temperature fluctuations $\delta T({\bm x}) = T({\bm x}) - T_{\text{eq}} $,
with $T({\bm x})$ the local temperature, have the form \cite{Landau_Lifshitz_V_1980}
\bse
\label{eqs:1.1}
\be
\langle\delta{T}({\bm x})\delta{T}({\bm x}')\rangle = \frac{T_{\text{eq}}^2}{c_V}\,\delta({\bm x} - {\bm x}')\ .
\label{eq:1.1a}
\ee
Here $\langle\ldots\rangle$ denotes an equilibrium statistical mechanics average, and $c_V$ is the specific heat per volume at constant volume. 
In wave-vector space, the same result is
\be
\langle\delta T({\bm k}_1)\delta T({\bm k}_2)\rangle = V\delta_{{\bm k}_1,-{\bm k}_2}\frac{ T_{\text{eq}}^2}{c_V}\ ,
\label{eq:1.1b}
\ee
\ese
where $V$ is the system volume. 

In a non-equilibrium steady state (NESS) characterized by a constant temperature gradient ${\bm\nabla}T$
the nature of this correlation changes drastically. There is a non-equilibrium contribution quadratic in ${\bm\nabla} T$
that diverges as $1/k^4$ in the limit of small wave numbers $k = \vert{\bm k}\vert$,\cite{Kirkpatrick_Cohen_Dorfman_1982c, 
Dorfman_Kirkpatrick_Sengers_1994, Ortiz_Sengers_2007}
\be
\frac{1}{V}\,\langle\delta T({\bm k})\delta T(-{\bm k})\rangle = \frac{T^2}{c_V} + \frac{ ({\hat{\bm k}}_{\perp}\!\!\cdot\!{\bm\nabla}T)^2\, T}{\rho\, D_T(\nu+D_T)\,k^4}\ .
\label{eq:1.2}
\ee
Here $T$ is the spatially averaged temperature, and $\rho$, $D_T$, and $\nu$ are the spatially averaged mass density, thermal
diffusion coefficient, and kinematic viscosity, respectively, of the fluid.\cite{approximations_footnote}
${\hat{\bm k}}_{\perp}$ is the unit wave vector perpendicular
to ${\bm k}$ in the plane spanned by ${\bm k}$ and ${\bm\nabla} T$. In the configuration sketched in Fig.~\ref{fig:1}, $T = (T_2+T_1)/2$, and
${\bm\nabla}T = {\hat z}\,\partial_z T$ with $\partial_z T = (T_2 - T_1)/L$ and ${\hat z}$ the unit vector in the $z$-direction, so
$({\hat{\bm k}}_{\perp}\!\cdot\!{\bm\nabla}T)^2 = (\partial_z T)^2 (k_x^2+k_y^2)/k^2$, see Eq.~(\ref{eq:A.2a}).{\cite{kperp_footnote}
\begin{figure}[b]
\includegraphics[width=8.5cm]{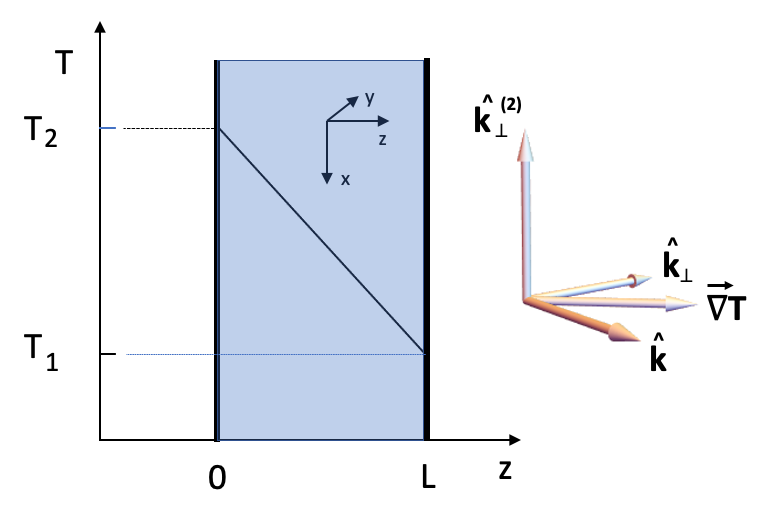}
\caption{A fluid subject to a constant temperature gradient in the $z$-direction between two parallel confining plates a distance $L$ apart.
              The three vectors $\hat{\bm k}$, $\hat{\bm k}_{\perp}$, and $\hat{\bm k}_{\perp}^{(2)}$ form a right-handed system that spans the wave-vector space.}
\label{fig:1}
\end{figure}

We note at this point that an arguably more physical choice of fluctuations than $\delta T$ to discuss in the presence of a 
temperature gradient would be the fluctuations of the entropy per particle, $\delta(S/N)$, which constitute the hydrodynamic 
heat mode in both classical\cite{Forster_1975} and quantum\cite{Belitz_Kirkpatrick_2022} fluids. However, in the absence of
pressure fluctuations $\delta(S/N)$ is proportional to $\delta T$, and historically the correlations of $\delta T$ have been
considered. See Appendix~\ref{app:A} for the relation between these fluctuations. 

The remarkably strong singularity expressed by Eq.~(\ref{eq:1.2}) has been derived theoretically using various techniques, including
kinetic theory,\cite{Kirkpatrick_Cohen_Dorfman_1982a, Kirkpatrick_Cohen_Dorfman_1982c}  
mode-coupling theory,\cite{Kirkpatrick_Cohen_Dorfman_1982a, Kirkpatrick_Cohen_Dorfman_1982c}  
and fluctuating hydrodynamics.\cite{Ronis_Procaccia_1982, Ortiz_Sengers_2007} This effect, as well as closely related ones,
have been observed by many experiments, see Ref.~\onlinecite{Sengers_Ortiz_Kirkpatrick_2016} and references therein. 
It has various physical consequences and interpretations. In
real space, it implies that correlations in a bulk fluid scale with the linear system size $L$
and decay on the same scale, see the discussion in Ref.~\onlinecite{Ortiz_Sengers_2007}. This in turn implies a generalized rigidity
of the fluid that is reminiscent of the effects of a spontaneously broken continuous symmetry in an equilibrium system, where Goldstone
modes lead to long-ranged correlations.\cite{Anderson_1984} By contrast, in a NESS rigidity is present even in the absence of any
broken symmetries. This has been discussed in Ref.~\onlinecite{Kirkpatrick_Belitz_Dorfman_2021}. This reference also showed that the second
spatial moment of a localized temperature perturbation that is accompanied by a perturbation of the shear velocity spreads ballistically, 
rather than diffusively, as a result of the generalized rigidity. That is, the root-mean-square displacement of a temperature perturbation
grows linearly with time for long times $t$, rather than as $t^{1/2}$ as for a diffusive process. 
The time scale associated with this ballistic spread is inversely proportional to
the product of the temperature gradient and the initial shear velocity.\cite{solid_footnote}

The goal of the current paper is two-fold. First, we will explore the question of how these effects manifest themselves 
in a quantum fluid. Second, we will show, for both quantum and classical fluids, that the long-ranged correlations can 
be probed via the system's response to macroscopic external perturbations, even though the usual fluctuation-dissipation 
theorem does not hold in the NESS. This is of particular interest at low temperatures, where fluctuations become small, 
or in systems where the non-equilibrium correlations dominate over the equilibrium contributions only at very small wave
numbers.

For the first goal, we will restrict ourselves to fermionic quantum fluids, but we note that analogous effects must be present 
in bosonic fluids,\cite{BEC_footnote} as the equations of hydrodynamics apply to those as well, see, e.g., 
Ref.~\onlinecite{Dorfman_vanBeijeren_Kirkpatrick_2021}. There are several crucial differences between the classical and
quantum cases. First, in the quantum regime one needs to distinguish between symmetrized, or anticommutator,
and antisymmetrized, or commutator, time correlation functions. If $\hat A$ and 
$\hat B$ are operators that correspond to two observables, then these two types of correlation functions are defined as
\bse
\label{eqs:1.3}
\bea
\Ssym_{AB}({\bm x},{\bm x}';t-t') &=& \frac{1}{2}\langle[\delta{\hat A}({\bm x},t),\delta{\hat B}({\bm x}',t')]_+\rangle\ ,\qquad
\label{eq:1.3a}\\
\chi_{AB}''({\bm x},{\bm x}';t-t') &=& \frac{1}{2\hbar}\langle[{\hat A}({\bm x},t),{\hat B}({\bm x}',t')]_-\rangle\ .\qquad
\label{eq:1.3b}
\eea
Here $[\ ,\ ]_{\mp}$ denotes a commutator or anticommutator, respectively, the average includes a quantum mechanical
average in addition to the statistical mechanics one, and $\delta{\hat A} = {\hat A} - \langle{\hat A}\rangle$. $\chi''$ is the 
customary notation for the commutator correlation function,\cite{Forster_1975}
with the double prime indicating that it is the spectrum, or spectral density, of a causal function.
$\Ssym_{AB}$ describes the correlations of spontaneous fluctuations. In equilibrium, $\chi_{AB}''$
describes the linear response of the system to external perturbations, and the temporal Fourier transforms
of the two correlation functions are related by\cite{Landau_Lifshitz_IX_1991, Forster_1975}
\be
\Ssym_{AB}({\bm x},{\bm x}';\omega) = \hbar\,\coth(\hbar\omega/2T)\,\chi_{AB}''({\bm x},{\bm x}';\omega)\ ,
\label{eq:1.3c}
\ee
\ese
which is a manifestation of the fluctuation-dissipation theorem.\cite{Nyquist_1928, Callen_Welton_1951}
For non-equilibrium systems Eq.~(\ref{eq:1.3c}) does not hold, no exact relation between the two correlation
functions is known, and $\chi_{AB}''$ in general is not a linear response function. Despite this lack of a
relation, one expects any long-range correlations that are present in $\Ssym$ to also be displayed in $\chi''$.
In particular, in a NESS $\Ssym_{TT}$ and $\chi''_{TT}$ should both contain the quantum analogs of the
long-range correlations present in Eq.~(\ref{eq:1.2}). Indeed, to leading order in the effects of the
temperature gradient Eq.~(\ref{eq:1.3c}) still holds with $T$ representing
the spatially averaged temperature.{\cite{approximations_footnote}} If a method for observing $\chi''_{TT}$ can be identified,
then this will provide an independent way of observing the long-range correlations. Identifying such a method
is our second goal. In Sec.~\ref{sec:IV} we will show that at least for the special case of the NESS considered here
the commutator correlations describe the bilinear response of the system to a field conjugate to the
shear velocity and is thus measurable via response or relaxation experiments.

In addition to these considerations, in a quantum fluid one needs to distinguish between the hydrodynamic regime
that is dominated by collisions between quasiparticles and the collisionless regime at asymptotically low temperatures where
all collision processes are frozen out.\cite{Pines_Nozieres_1989} 

For an educated guess of the results for the correlation functions in the quantum regime we recall the origin of the effect in the 
classical case. The crucial term, in this context, in the Navier-Stokes equations is the streaming term that couples the spatial 
temperature fluctuations to the fluid velocity ${\bm u}$ be means of a bilinear term ${\bm u}\cdot{\bm\nabla T}$. In an 
equilibrium system this term is quadratic in the fluctuations. However, in a NESS that is characterized by an externally fixed 
temperature gradient it is linear in the fluctuating quantity ${\bm u}$. The transverse, or shear, 
velocity is diffusive, and the solution of the coupled equations effectively results in a product of two diffusive contributions to
the temperature correlation function, each of which scales as $1/k^2$. In the quantum hydrodynamic regime the structure of the
equations of motion is the same as in the classical case,\cite{Kirkpatrick_Belitz_2022} which implies that the anticommutator,
or symmetrized, temperature correlation function will still diverge as $1/k^4$. The commutator, or antisymmetrized, correlation
function in the hydrodynamic regime, to leading order in the effects of the temperature gradient, is proportional to the 
anticommutator one, see Eq.~(\ref{eq:1.3c}) and the comments following that equation, and will also diverge as $1/k^4$. In the
collisionless regime there are two modifications. First, the relevant velocity mode is ballistic rather than diffusive, which
changes the $1/k^4$ divergence to $1/k^2$. Second, one needs to remember that the temperature prefactor in the
nonequilibrium term in Eq.~(\ref{eq:1.2}) is actually $(\hbar\omega/2)\,\coth(\hbar\omega/2T)$, which essentially is the greater of the 
temperature and the shear mode energy. In the part of the collisionless regime where the temperature is greater than the 
ballistic mode energy the anticommutator correlation function is thus expected to still show the $1/k^2$ behavior, whereas in the 
regime where the temperature is the smallest energy scale it should diverge as $1/k$. The commutator correlation function,
which to leading order in the effects of the temperature gradient is related to the anticommutator one by means of the
same factor, is expected to diverge as $1/k^2$ everywhere in the collisionless regime. Since these effects are due to the shear
velocity coupling to the temperature they will be present in both neutral and charged Fermi fluids, as the Coulomb
interaction affects only the longitudinal fluid velocity.

As we will show below, these expectations are borne out. For the technical derivation we will use a generalization of the 
fluctuating quantum kinetic theory developed in Ref.~\onlinecite{Kirkpatrick_Belitz_2022}. 

The organization of this paper is as follows. In Sec.~\ref{sec:II} we consider the nonlinear version of the fluctuating 
Landau-Boltzmann equation that was discussed in Ref.~\onlinecite{Kirkpatrick_Belitz_2022} 
and derive nonlinear fluctuating Navier-Stokes equations for a fermionic quantum fluid. In Sec.~\ref{sec:III} we simplify
these quantum kinetic theories and use them to describe a Fermi liquid subject to a fixed temperature gradient. In Sec.~\ref{sec:IV}
we establish a relation between the commutator correlation functions and the bilinear response of the fluid to external perturbations
and we discuss the anomalously fast propagation of a temperature perturbation in a NESS. We conclude
in Sec.~\ref{sec:V} with a discussion of our results. Various technical details are relegated to appendices. A brief account of
some of our results has previously been given in Ref.~\onlinecite{Kirkpatrick_Belitz_2023b}.

\section{Quantum Hydrodynamics}
\label{sec:II}

In this section we derive and discuss the relevant kinetic equations for our problem. Underlying all of them is the Boltzmann-Landau
kinetic equation for the single-particle phase space distribution function, or $\mu$-space distribution function in the terminology of
Ehrenfest,\cite{Ehrenfest_Ehrenfest_1911} from which one can derive Navier-Stokes equations by means of a
Chapman-Enskog expansion. The Navier-Stokes equations are valid in the hydrodynamic regime, which is dominated by collisions.
In order to study the collisionless regime one has to consider the underlying kinetic equation directly.\cite{Navier-Stokes_validity_footnote} In order to calculate 
dynamic correlation functions Langevin forces need to be added to all of these equations. 

\subsection{Kinetic Equations for Averaged Variables}
\label{subsec:II.A}

\subsubsection{The Boltzmann-Landau and Uehling-Uhlenbeck equations} 
% ch 154 $1
\label{subsubsec:II.A.1}

Consider the single-particle phase space or $\mu$-space spanned by the position ${\bm x}$ and the momentum ${\bm p}$
of a particle\cite{Ehrenfest_Ehrenfest_1911} (we will consider only fermions). Let ${\hat f}({\bm p},{\bm x},t)$ be the operator-valued single-particle distribution
function,\cite{caret_footnote} let $f_{\bm p}({\bm x},t) = \langle{\hat f}({\bm p},{\bm x},t)\rangle$ be its average, where $\langle \ldots\rangle$ denotes
a quantum mechanical expectation value plus a statistical mechanics average, and let $\delta f_{\bm p}({\bm x},t) = f_{\bm p}({\bm x},t) - f_{\bm p}^{\text eq}$
be its deviation from the equilibrium distribution. Further, let ${\hat\epsilon}({\bm p},{\bm x},t)$ be the single-particle energy,
and $\epsilon_{\bm p}({\bm x},t) = \langle {\hat\epsilon}({\bm p},{\bm x},t)\rangle$ its average. The time evolution of $f_{\bm p}$ is governed by the 
Boltzmann-Landau kinetic equation\cite{Landau_Lifshitz_X_1981}
\be
\partial_t f_{\bm p} + ({\bm\nabla_{\bm x}} f_{\bm p})\cdot{\bm\nabla}_{\bm p}\epsilon_{\bm p} - ({\bm\nabla}_{\bm p} f_{\bm p})\cdot{\bm\nabla}_{\bm x}\epsilon_{\bm p} = C(f)_{\bm p}\ .
\label{eq:2.1}
\ee
Here and in the remainder of this subsection we drop the real-space and time arguments as long as they are the same for all quantities in a given equation.
The terms on the left-hand side of Eq.~(\ref{eq:2.1}) represent the total time derivative $d f_{\bm p}/dt$, which is balanced by the
collision integral on the right-hand side, i.e., the temporal change of $f_{\bm p}$ due to collisions between quasiparticles. 
The latter is given by\cite{Landau_Lifshitz_X_1981}
\begin{widetext}
\bea
C(f)_{\bm p} &=& \frac{1}{V^3} \sum_{{\bm p}'\!,{\bm p}_1,{\bm p}_1'}  
     W({\bm p},{\bm p}_1;{\bm p}',{\bm p}_1')\,\delta(\epsilon_p + \epsilon_{p_1} - \epsilon_{p'} - \epsilon_{p_1'})\, \delta({\bm p}+{\bm p}_1-{\bm p}'-{\bm p}_1') 
\nonumber\\
&& \times\left[ f_{{\bm p}'} f_{{\bm p}_1'} (1 - f_{\bm p})(1 - f_{{\bm p}_1}) - f_{\bm p} f_{{\bm p}_1} (1 - f_{{\bm p}'})(1 - f_{{\bm p}_1'})\right]
\label{eq:2.2}
\eea
\end{widetext}
with $V$ the system volume. The transition rate $W$ is positive and has the symmetry properties
\bse
\label{eqs:2.3}
\bea
W({\bm p},{\bm p}_1;{\bm p}',{\bm p}_1') &=& W(-{\bm p},-{\bm p}_1;-{\bm p}',-{\bm p}_1')
\label{eq:2.3a}\\
&=& W({\bm p}',{\bm p}_1';{\bm p},{\bm p}_1)
\label{eq:2.3b}\\
&=& W({\bm p}_1,{\bm p};{\bm p}_1',{\bm p}')\ ,
\label{eq:2.3c}
\eea
\ese
which express invariance under spatial inversions, time reversal, and interchange of particles.

In order to fully define Eq.~(\ref{eq:2.1}) we also need to specify $\epsilon_{\bm p}$. Within Landau Fermi-liquid (LFL)
theory one has\cite{Landau_Lifshitz_IX_1991}
\be
\epsilon_{\bm p}({\bm x},t) = \epsilon_p + \frac{1}{V}\sum_{{\bm p}'} F({\bm p},{\bm p}')\,\delta f_{{\bm p}'}({\bm x},t)
\label{eq:2.4}
\ee
where $\epsilon_p = p^2/2m$, with $m$ the bare fermion mass, is the equilibrium single-particle energy\cite{single-particle_energy_footnote} 
and $F({\bm p},{\bm p}')$ is Landau's interaction function.
At zero temperature it can be expanded in spherical harmonics on the Fermi surface and parameterized in terms of
the LFL parameters $F_0$, $F_1$, etc. LFL theory relies on linear variational arguments and in general is compatible
only with the linearized version of the kinetic equation (\ref{eq:2.1}). If one uses the full nonlinear equation one obtains
a consistent theory only if one replaces the function $F({\bm p},{\bm p}')$ by a constant. This is tantamount to keeping
only the Landau parameter $F_0$ and is equivalent to treating the quasiparticle interaction in Hartree-Fock approximation.
For our purposes the interaction makes no qualitative difference since it does not change the nature of the crucial soft
modes. For our explicit calculation we will therefore ignore the interaction altogether, which amounts to dropping the last 
term on the left-hand side of Eq.~(\ref{eq:2.1}) and replacing ${\bm\nabla}_{\bm p}\epsilon_{\bm p}$ by 
${\bm p}/m \equiv {\bm v}_{\bm p}$. We then have
\be
\partial_t f_{\bm p} + {\bm v}_{\bm p}\cdot({\bm\nabla_{\bm x}} f_{\bm p}) = C(f)_{\bm p}\ .
\label{eq:2.5}
\ee
In this form the kinetic equation is known as the Uehling-Uhlenbeck 
equation.\cite{Uehling_Uhlenbeck_1933, Dorfman_vanBeijeren_Kirkpatrick_2021, external_force_footnote}
Consistent with these simplifications, we ignore the spin degree of freedom.

\subsubsection{The local Fermi-Dirac distribution}
\label{subsubsec:II.A.2}

The Boltzmann-Landau equation (\ref{eq:2.1}) allows for an $H$-theorem in analogy to the classical Boltzmann
equation, see, e.g., ch. 10.3.5. in Ref.~\onlinecite{Dorfman_vanBeijeren_Kirkpatrick_2021}, or Appendix D in 
Ref.~\onlinecite{Kirkpatrick_Belitz_2022}. The $H$-theorem shows that the
entropy change as a function of time is positive semi-definite, and zero if and only if the distribution $f_{\bm p}$
is equal to the equilibrium Fermi-Dirac distribution
\be
f_{\bm p}^{\text{eq}} = \frac{1}{e^{(\epsilon_p - \mu)/T}+1}\ ,
\label{eq:2.6}
\ee
with $\mu$ the chemical potential. (We use units such that $\kB = 1$. We will also put $\hbar=1$; however, see 
Ref.~\onlinecite{hbar_footnote}.) This form of the equilibrium distribution results from the fact that the entropy
production vanishes if and only if $\log(f_{\bm p}/(1-f_{\bm p}))$ is a collision invariant and hence a linear
combination of the five basic collision invariants, viz., particle number, momentum, and energy.
The equilibrium distribution solves the kinetic equation since it is independent
of space and time and has the property $C(f^{\text{eq}}) = 0$. The latter property remains true if we consider a
local Fermi-Dirac distribution
\bse
\label{eqs:2.7}
\be
f_{\bm p}^{(0)}({\bm x},t) = \frac{1}{\exp\left[\left(\frac{({\bm p} - m{\bm u}({\bm x},t))^2}{2m} - \mu({\bm x},t)\right)/T({\bm x},t)\right] + 1}\ .
\label{eq:2.7a}
\ee
Note that $f_{\bm p}^{(0)}$ is not a solution of either Eq.~(\ref{eq:2.1}) or (\ref{eq:2.5}). It satisfies $C(f^{(0)}) = 0$ for
arbitrary functions ${\bm u}$, $\mu$, and $T$, but the concept of a local equilibrium distribution is most useful
if one chooses these functions to be the physical fluid velocity, chemical potential, and temperature, respectively.
That is, we require
\be
\frac{1}{V}\sum_{\bm p} f_{\bm p}^{(0)}({\bm x},t) = n({\bm x},t)
\label{eq:2.7b}
\ee
with $n({\bm x},t)$ the physical number density, and 
\be
\frac{1}{V}\sum_{\bm p} {\bm p}\, f_{\bm p}^{(0)}({\bm x},t) = \rho({\bm x},t)\,{\bm u}({\bm x},t)\ ,
\label{eq:2.7c}
\ee
\ese
with $\rho({\bm x},t) = m\,n({\bm x},t)$ the physical mass density.

\subsubsection{Balance equations}
\label{subsubsec:II.A.3}

The kinetic equation yields balance equations, in complete analogy to the classical case, for the five collision invariants: particle number or mass, 
momentum, and energy.\cite{Cercignani_1988, Ortiz_Sengers_2007, Dorfman_vanBeijeren_Kirkpatrick_2021}
For a discussion of these balance laws in terms of conservation laws, the fluid flow, and forces acting on a volume element
in the fluid, see Ref.~\onlinecite{Ortiz_Sengers_2007}.

\smallskip
\paragraph{Mass balance}
\label{par:II.A.3a}

By summing Eq.~(\ref{eq:2.5}) over the momentum ${\bm p}$ we obtain the mass balance equation
\be
\partial_t\, \rho({\bm x},t) + {\bm\nabla}\cdot\left(\rho({\bm x},t){\bm u}({\bm x},t)\right) = 0
\label{eq:2.8}
\ee
which expresses the local conservation of mass. 

\smallskip
\paragraph{Momentum balance}
\label{par:II.A.3b}

By multiplying Eq.~(\ref{eq:2.5}) with a component $p_i$ of the momentum and summing over ${\bm p}$ we obtain
the velocity equation
\bse
\label{eqs:2.9}
\be
\partial_t u_i({\bm x},t) + \left({\bm u}({\bm x},t)\cdot{\bm\nabla}\right) u_i({\bm x},t) = \frac{-1}{\rho({\bm x},t)}\,\partial_j P^{ij}({\bm x},t)\ ,
\label{eq:2.9a}
\ee
with
\be
P^{ij}({\bm x},t) = \frac{m}{V}\sum_{\bm p} \left(v_{\bm p}^i - u^i({\bm x},t)\right) \left(v_{\bm p}^j - u^j({\bm x},t)\right) f_{\bm p}({\bm x},t)
\label{eq:2.9b}
\ee
\ese
the kinetic part of the pressure tensor. Here $\partial_j \equiv \partial/\partial x^j$, and summation over repeated indices is implied.

\smallskip
\paragraph{Energy balance}
\label{par:II.A.3c}
Finally, by multiplying with $m\left({\bm v}_{\bm p} - {\bm u}({\bm x},t)\right)^2/2$ and summing over ${\bm p}$ we obtain
a balance equation for the kinetic energy density
\be
e({\bm x},t) = \frac{m}{2}\, \frac{1}{V}\sum_{\bm p} \left({\bm v}_{\bm p} - {\bm u}({\bm x},t)\right)^2 f_{\bm p}({\bm x},t)
\label{eq:2.10}
\ee
in the form
\bse
\label{eqs:2.11}
\bea
\partial_t\,e({\bm x},t) + {\bm u}({\bm x},t)\cdot{\bm\nabla} e({\bm x},t) &=& - e({\bm x},t){\bm\nabla}\cdot{\bm u}({\bm x},t)
\nonumber\\
&& \hskip -100pt  - {\bm\nabla}\cdot{\bm j}_e({\bm x},t)
     - P^{ij}({\bm x},t)\partial_i u_j({\bm x},t)\ ,
\label{eq:2.11a}
\eea
with
\be
{\bm j}_e({\bm x},t) =  \frac{m}{2}\, \frac{1}{V}\sum_{\bm p} \left({\bm v}_{\bm p} - {\bm u}({\bm x},t)\right) \left({\bm v}_{\bm p} - {\bm u}({\bm x},t)\right)^2
     f_{\bm p}({\bm x},t)
\label{eq:2.11b}
\ee
\ese
the kinetic energy current density or heat flux.

It is useful to rewrite the energy balance equation as an equation for the temperature $T$. To this end we consider $T$ a function
of $e$ and $n$. Then variations of these three quantities are related by
\be
\delta T = \left(\frac{\partial T}{\partial e}\right)_{N,V} \delta e + \left(\frac{\partial T}{\partial n}\right)_{E,V} \delta n\ .
\label{eq:2.12}
\ee
But $(\partial T/\partial e)_{N,V} = 1/c_V$, and general thermodynamic identities yield\cite{thermodynamics_footnote}
\be
c_V  \left(\frac{\partial T}{\partial n}\right)_{E,V} = -\mu + T  \left(\frac{\partial \mu}{\partial T}\right)_{N,V} \equiv -\tilde\mu\ .
\label{eq:2.13}
\ee
We thus have
\be
c_V \delta T({\bm x},t) = \delta e({\bm x},t) - \tilde\mu\, \delta n({\bm x},t)\ .
\label{eq:2.14}
\ee
Together with the mass balance equation (\ref{eq:2.8}) this yields
\bea
\left(\partial_t + {\bm u}({\bm x},t)\!\cdot\!{\bm\nabla}\right) e({\bm x},t) &=& c_V \left(\partial_t + {\bm u}({\bm x},t)\!\cdot\!{\bm\nabla}\right) T({\bm x},t)\nonumber\\  
&& - \tilde\mu\, n({\bm x},t){\bm\nabla}\!\cdot\!{\bm u}({\bm x},t)\ ,
\label{eq:2.15}
\eea
which can be used to rewrite Eq.~(\ref{eq:2.11a}) as an equation for $T$ instead of $e$. Note that all thermodynamic
derivatives, and in particular $\tilde\mu$, are in principle space and time dependent. If we replace the derivatives by
their average values,\cite{approximations_footnote} as is usually done in nonlinear hydrodynamics,\cite{Ortiz_Sengers_2007} we can make use
of another thermodynamic identity that relates $\tilde\mu$ to the average energy density $e$ and pressure $p$ and another derivative:
\be
n\,\tilde\mu = e + p - T(\partial p/\partial T)_{N,V}
\label{eq:2.16}
\ee
Finally, we drop all other nonlinearities except for the crucial coupling between the temperature fluctuations and the
fluid velocity. We then obtain the temperature equation in the form
\bea
c_V \left(\partial_t + {\bm u}({\bm x},t)\!\cdot\!{\bm\nabla}\right) T({\bm x},t) &=& - {\bm\nabla}\!\cdot\!{\bm j}_e({\bm x},t)
\nonumber\\
 && \hskip -75pt - T \left(\frac{\partial p}{\partial T}\right)_{N,V} {\bm\nabla}\!\cdot\!{\bm u}({\bm x},t) 
\label{eq:2.17}
\eea
where $T$ in the second term on the right-hand side is the average temperature.

\subsection{Navier-Stokes equations}
\label{subsec:II.B}

The Navier-Stokes equations can be derived from very general arguments, and hence clearly are valid
for quantum fluids as well as for classical ones. However, for completeness we derive them in this section
from the quantum kinetic equation. As in the classical case, the Navier-Stokes equations are actually more generally
valid then the derivation suggests, see Appendix~\ref{app:D}.

\subsubsection{Chapman-Enskog expansion}
\label{subsubsec:II.C.1}

To derive closed fluid-dynamics equations we employ the Chapman-Enskog method in the same way as in
classical fluids.\cite{Chapman_Cowling_1970, Dorfman_vanBeijeren_Kirkpatrick_2021} The basic idea is
to introduce a small parameter $\alpha = O(\ell/L)$ on the order of the ratio of the mean-free
path between collision, $\ell$, and a macroscopic lenth $L$ that scales as the inverse spatial gradient in
the kinetic equation. This small parameter does not appear
explicitly in the kinetic equation; rather, it is introduced by hand by multiplying the right-hand side of the kinetic
equation (\ref{eq:2.1}) by $1/\alpha$,
\be
\partial_t f_{\bm p} + ({\bm\nabla_{\bm x}} f_{\bm p})\cdot{\bm\nabla}_{\bm p}\epsilon_{\bm p} - ({\bm\nabla}_{\bm p} f_{\bm p})\cdot{\bm\nabla}_{\bm x}\epsilon_{\bm p} = \frac{1}{\alpha}\,C(f)\ .
\tag{2.1'}
\ee
Expanding $f_{\bm p}$ in powers of $\alpha$,
\be
f_{\bm p} = f_{\bm p}^{(0)} + \alpha f_{\bm p}^{(1)} + O(\alpha^2)\ ,
\label{eq:2.18}
\ee
yields a hierarchy of equations for the $f_{\bm p}^{(n)}$, order by order in $\alpha$. After truncating the expansion
at the desired order one puts $\alpha=1$. 

\subsubsection{Euler Equations}
\label{subsubsec:II.C.2}

To zeroth order in the Chapman-Enskog expansion we have
\be
C(f_{\bm p}^{(0)}) = 0\ .
\label{eq:2.19}
\ee
The solution of this equation is not unique: as we saw in Sec.~\ref{subsubsec:II.A.2}, both the global and the 
local equilibrium distributions satisfy Eq.~(\ref{eq:2.19}). Following the usual procedure in the classical case, we choose
the latter.\cite{CE_footnote} From Eq.~(\ref{eq:2.9b}) we see that in this approximation the pressure tensor is
diagonal  and given by
\bse
\label{eqs:2.20}
\be
P_{ij}({\bm x},t) \approx P_{ij}^{(0)} = \delta_{ij}\, p({\bm x},t)
\label{eq:2.20a}
\ee
with
\be
p({\bm x},t) = \frac{2}{3}\,e({\bm x},t)
\label{eq:2.20b}
\ee
\ese
the hydrostatic pressure. Note that this is the exact relation between the pressure and the energy for an
ideal Fermi gas (or any nonrelativistic ideal gas). This specifies the right-hand side of the momentum balance equation (\ref{eq:2.9a}). For
the heat flux we obtain from Eq.~(\ref{eq:2.11b})
\be
{\bm j}_e({\bm x},t) \approx 0\ .
\label{eq:2.21}
\ee

For the hydrodynamic equations at this order in the Chapman-Enskog expansion we thus obtain Euler's
equations, viz.: The mass equation as given by Eq.~(\ref{eq:2.8}), the momentum equation reads
\be
\partial_t {\bm u}({\bm x},t) + \left({\bm u}({\bm x},t)\cdot{\bm\nabla}\right){\bm u}({\bm x},t) = \frac{-1}{\rho({\bm x},t)}\,{\bm\nabla}p({\bm x},t)\ ,
\label{eq:2.22}
\ee
and the energy equation is
\be
\partial_t e({\bm x},t) + \left({\bm u}({\bm x},t)\cdot{\bm\nabla}\right)e({\bm x},t) = p({\bm x},t)\,{\bm\nabla}\cdot{\bm u}({\bm x},t)\ .
\label{eq:2.23}
\ee
Alternatively, we can write the energy equation as an equation for the temperature. From Eq.~(\ref{eq:2.17}) we have
\bea
c_V \partial_t T({\bm x},t) + c_V \left({\bm u}({\bm x},t)\cdot{\bm\nabla}\right)T({\bm x},t) 
                        &=& \qquad\qquad
                        \nonumber\\
                        && \hskip -100pt -T \left(\frac{\partial p}{\partial T}\right)_{N,V} {\bm\nabla}\cdot{\bm u}({\bm x},t)\ .
\label{eq:2.24}
\eea
Note that the energy equation in the form of (\ref{eq:2.23}) is exact to this order, whereas in the temperature equation (\ref{eq:2.24})
the thermodynamic derivatives have been replaced by their average values, and this includes the $T$ prefactor on the right-hand 
side.\cite{approximations_footnote}

In addition to ${\bm u}({\bm x},t)$ and $T({\bm x},t)$, which are governed by the Euler equations, $f_{\bm p}^{(0)}$ depends on
$\mu({\bm x},t)$, which is given implicitly by the requirement (\ref{eq:2.7b}).      

\subsubsection{Navier-Stokes equations}
\label{subsubsec:II.B.3}           

To first order in the expansion in powers of $\alpha$ we have
\be
(\partial_t + {\bm v}_{\bm p}\cdot{\bm\nabla_{\bm x}})f_{\bm p}^{(0)}({\bm x},t) = \Lambda({\bm p})\,f_{\bm p}^{(1)}({\bm x},t)\ ,
\label{eq:2.25}
\ee
with $\Lambda({\bm p})$ a linearized collision operator that is given by $C(f)_{\bm p}$ expanded to linear order in $f_{\bm p}^{(1)}$. 
Note that the third term on the right-hand side of Eq.~(2.1'), which is omitted in the Uehling-Uhlenbeck equation, does not
contribute to this order. 

\paragraph{$\mu$-space distribution to first order}
\label{par:II.B.3a}

$f_{\bm p}^{(0)}$ depends on ${\bm x}$ and $t$ through ${\bm u}({\bm x},t)$ and $T({\bm x},t)$, as well as $\mu({\bm x},t)$,
which in turn depends on ${\bm u}$ and $T$ through Eq.~(\ref{eq:2.7b}). By calculating the derivatives of $f_{\bm p}^{(0)}$,
Eq.~(\ref{eq:2.7a}), with respect to ${\bm u}$, $T$, and $\mu$, and using the Euler equations, we obtain a linear integral
equation for $f_{\bm p}^{(1)}$. If we define
\begin{widetext}
\bse
\label{eqs:2.26}
\be
{\bm c}({\bm x},t) = {\bm p}/m - {\bm u}({\bm x},t)
\label{eq:2.26a}
\ee
the latter can be written
\bea
\Lambda({\bm p})\,f_{\bm p}^{(1)} &=& f_{\bm p}^{(0)} \left(1 - f_{\bm p}^{(0)}\right)\Biggl\{\frac{m}{T}\,c_i c_j \partial^j u^i - \frac{1}{c_V}\left(\frac{\partial p}{\partial T}\right)_{N,V} \left[\frac{1}{T}\left(\frac{m}{2}\,{\bm c}^2 - \mu\right) + \left(\frac{\partial\mu}{\partial T}\right)_{N,V}\right] \partial_i u^i - \frac{n}{T(\partial n/\partial\mu)_{T,V}}\,\partial^i u_i
\nonumber\\
&& \hskip 70pt + \frac{1}{T} \left[\frac{-1}{n}\left(\frac{\partial p}{\partial T}\right)_{N,V} + \frac{1}{T}\left(\frac{m}{2}\,{\bm c}^2 - \mu\right) + \left(\frac{\partial\mu}{\partial T}\right)_{N,V}\right]({\bm c}\cdot{\bm\nabla})T\Biggr\}\ ,
\label{eq:2.26b}
\eea
\ese
where we have omitted the obvious dependences on space and time. In classical kinetic theory the equivalent of the relative velocity ${\bm c}$ is sometimes
called `peculiar velocity'.

\paragraph{Pressure tensor and heat flux to first order}
\label{par:II.B.3b}

The terms on the right-hand side of Eq.~(\ref{eq:2.26b}) that are even in ${\bm c}$
yield the first-order contribution to the pressure tensor via Eq.~(\ref{eq:2.9b}). We find
\be
P_{ij}^{(1)} = \frac{m^2}{2 T}\, \left(\partial^k u^l + \partial^l u^k\right) \frac{1}{V} \sum_{\bm p} c_i c_j \Lambda^{-1}({\bm p}) \left(c_k c_l - \frac{1}{3}\,\delta_{kl} {\bm c}^2\right)
     f_{\bm p}^{(0)} \left(1 - f_{\bm p}^{(0)}\right)\ .
\label{eq:2.27}
\ee
\end{widetext}
Here we have used Eq.~(\ref{eq:2.20b}) as well as the identity 
\be
\mu - T(\partial\mu/\partial T)_{N,V} - \frac{3}{2}\,n(\partial \mu/\partial n)_{T,V} = 0\ ,
\label{eq:2.28}
\ee
which follows from the fact that the chemical potential as a function of $n$ and $T$ has the form $\mu = n^{2/3} f_{\mu}(T^2/n^{4/3})$, with $f_{\mu}$
a scaling function. Equation~(\ref{eq:2.27}) can be cast in a more standard form by realizing that $P_{ij}^{(1)}$ is a traceless symmetric tensor that
is linear in the symmetric tensor $\partial_i u_j + \partial_j u_i$ and therefore must be proportional to the traceless version of the latter. Restoring
the dependence on space and time we find, to first order in the Chapman-Enskog expansion,
\bse
\label{eqs:2.29}
\bea
P_{ij}({\bm x},t) &=& \delta_{ij} p({\bm x},t) - \eta \biggl[\partial_i u_j({\bm x},t) + \partial_j u_i({\bm x},t) 
\nonumber\\
     && \hskip 50pt \left. - \frac{2}{3}\,\delta_{ij}\,{\bm\nabla}\cdot{\bm u}({\bm x},t)\right]
\label{eq:2.29a}
\eea
where the shear viscosity $\eta$ is given by
\be
\eta = \frac{-m^2}{T}\,\frac{1}{V}\sum_{\bm p} c_1 c_2\, \Lambda^{-1}({\bm p}) \, c_1 c_2\,f_{\bm p}^{(0)}\left(1 - f_{\bm p}^{(0)}\right)\ .
\label{eq:2.29b}
\ee
\ese
Note that $f_{\bm p}^{(0)}$, and hence also $\Lambda$, depend on ${\bm p}$ only via the combination ${\bm p}/m - {\bm u} = {\bm c}$,
so the space-time dependence of $\bm c$ drops out via the sum over ${\bm p}$. However, $\eta$ does depend on ${\bm x}$ and $t$ via $\mu$
and $T$; we consider $\eta$ in Eq.~(\ref{eq:2.29a}) the averaged value.\cite{approximations_footnote}

The terms on the right-hand side of Eq.~(\ref{eq:2.26b}) that are odd in ${\bm c}$ determine the heat flux to first order in the
Chapman-Enskog expansion. Equation~(\ref{eq:2.11b}) yields
\be
{\bm j}_e({\bm x},t) = -\kappa\,{\bm\nabla} T({\bm x},t)
\label{eq:2.30}
\ee
with the thermal conductivity $\kappa$ given by 
\be
\kappa = \frac{-m}{3 T^2}\,\frac{1}{V} \sum_{\bm p} c\,\frac{m}{2}\,{\bm c}^2\,\Lambda^{-1}({\bm p})\,c\,\psi_h({\bm c})\,f_{\bm p}^{(0)}\left(1 - f_{\bm p}^{(0)}\right)\ .
\label{eq:2.31}
\ee
Here $c = \vert{\bm c}\vert$, and 
\be
\psi_h({\bm c}) %&=& \frac{m}{2}\,{\bm c}^2 - \mu + T\left(\frac{\partial\mu}{\partial T}\right)_{N,V} - \frac{T}{n}\left(\frac{\partial p}{\partial T}\right)_{N,V}
%\nonumber\\
= \frac{m}{2}\,{\bm c}^2 - \mu - \frac{Ts}{n}\ ,
\label{eq:2.32}
\ee
with $s$ the entropy per volume, is the heat mode.\cite{heat_mode_footnote} The asymmetric form of the integrand in Eq.~(\ref{eq:2.31}) is seemingly at odds with the
Kubo formula for the thermal conductivity. The resolution of this problem is the observation that Eqs.~(\ref{eq:2.7b}, \ref{eq:2.7c}) imply
$\sum_{\bm p}{\bm c}\,f_{\bm p}^{(1)} = 0$. One can therefore add an arbitrary term independent of $\bm c$ to the factor of
$m{\bm c}^2/2$ in Eq.~({\ref{eq:2.31}), and this allows us to rewrite the expression for $\kappa$ in the symmetric form
\be
\kappa = \frac{-m}{3T^2}\,\frac{1}{V} \sum_{\bm p} c \,\psi_h({\bm c})\,\Lambda^{-1}({\bm p})\,c\,\psi_h({\bm c})\,f_{\bm p}^{(0)}\left(1 - f_{\bm p}^{(0)}\right)\ .
\tag{2.31'}
\ee
This argument leading to the symmetric expression for $\kappa$ is the same as in the classical case.\cite{Dorfman_vanBeijeren_Kirkpatrick_2021}

\paragraph{Navier-Stokes equations}
\label{par:II.B.3c}

We are now in a position to assemble the hydrodynamic equations to first order in the Chapman-Enskog expansion. 
The mass equation is still given by Eq.~(\ref{eq:2.8}), which is exact. For the velocity equation we have, from Eqs.~(\ref{eq:2.9a})
and (\ref{eq:2.29a}),
\begin{widetext}
\be
\partial_t u_i({\bm x},t) + u_j({\bm x},t) \partial_j u_i({\bm x},t) = \frac{-1}{\rho({\bm x},t)}\,\partial_i\, p({\bm x},t) 
     + \frac{\eta}{\rho({\bm x},t)}\,\partial_j  \left[\partial_i u_j({\bm x},t) + \partial_j u_i({\bm x},t) - \frac{2}{3}\,\delta_{ij}\,{\bm\nabla}\cdot{\bm u}({\bm x},t)\right]\ ,
\label{eq:2.33}
\ee
with $p({\bm x},t)$ the hydrostatic pressure from Eq.~(\ref{eq:2.20b}) and $\eta$ the shear viscosity from Eq.~(\ref{eq:2.29b}). 
Finally, for the heat equation we obtain, from Eqs.~(\ref{eq:2.17}) and (\ref{eq:2.30}),
\be
\partial_t T({\bm x},t) + {\bm u}({\bm x},t)\cdot{\bm\nabla} T({\bm x},t) = \frac{-T}{c_V}\left(\frac{\partial p}{\partial T}\right)_{N,V} {\bm\nabla}\cdot{\bm u}({\bm x},t)
     + \frac{\kappa}{c_V}\,{\bm\nabla}^2 T({\bm x},t)\ .
\label{eq:2.34}
\ee
\end{widetext}
Equations (\ref{eq:2.8}), (\ref{eq:2.33}), and (\ref{eq:2.34}) are simplified versions of the standard Navier-Stokes equations familiar
from classical hydrodynamics;\cite{Forster_1975, Chaikin_Lubensky_1995, Navier_Stokes_footnote} the fermionic nature of the fluid is reflected only in the
explicit expressions for the transport coefficients $\eta$ and $\kappa$. This was to be expected, since the behavior of a fluid in the
hydrodynamic regime depends only on very general physical principles that are independent of the microscopic nature of
the fluid. We note again that we have replaced the transport coefficients $\eta$ and $\kappa$, as well as the thermodynamic
derivatives in Eqs.~(\ref{eq:2.33}) and (\ref{eq:2.34}), by their average values.\cite{approximations_footnote}

\subsection{Fluctuating Quantum Navier-Stokes equations}
\label{subsec:II.C}

The Navier-Stokes equations contain many nonlinearities that make them notoriously hard to solve. For fluctuations about
an equilibrium state, the bilinear ${\bm u}\cdot{\bm\nabla}T$ term in Eq.~(\ref{eq:2.34}) is one of these nonlinearities. In a
NESS characterized by a constant temperature gradient, ${\bm\nabla} T$ is no longer a fluctuation, and the leading
contribution to this term is linear in the small fluctuation ${\bm u}$. We can thus linearize the theory by replacing 
${\bm\nabla} T({\bm x},t)$ in the ${\bm u}\cdot{\bm\nabla} T$ term with the externally fixed temperature gradient ${\bm\nabla} T$, 
which makes this term linear, and also dropping all other nonlinearities. Furthermore, for our purposes we are not interested in the 
coupling of the temperature gradient to sound waves, which occur on a time scale that is much faster than the slow fluctuations
of the transverse fluid velocity, whose dynamics are diffusive. The sound modes are linear combinations of fluctuations of the
longitudinal part of the fluid velocity ${\bm u}$ and pressure fluctuations, see Eq.~(\ref{eq:A.5}), or Eq.~(3.25) in 
Ref.~\onlinecite{Belitz_Kirkpatrick_2022}. Accordingly, we work at constant pressure and keep only the diffusive transverse 
components ${\bm u}_{\perp}$ of the fluid velocity in the ${\bm u}\cdot{\bm\nabla}T$ coupling term in Eq.~(\ref{eq:2.34}). 
For the latter, the linearized Eq.~(\ref{eq:2.33}) simplifies to a diffusion equation
\bse
\label{eqs:2.35}
\be
\partial_t {\bm u}_{\perp}({\bm x},t) = \nu {\bm\nabla}^2 {\bm u}_{\perp}({\bm x},t)\ ,
\label{eq:2.35a}
\ee
with $\nu = \eta/\rho$ the kinematic viscosity. The longitudinal part of the fluid velocity scales linearly with the wave number,
and hence the ${\bm\nabla}\cdot{\bm u}$ term on the right-hand side of Eq.~(\ref{eq:2.34}) is of the same order in a gradient
expansion as the ${\bm\nabla}^2 T$ term. At constant pressure it effectively turns the 
$\kappa/c_V$ coefficient of the ${\bm\nabla}^2 T$ term into $\kappa/c_p$, and we find
\be
\partial_t T({\bm x},t) + {\bm u}_{\perp}({\bm x},t)\cdot{\bm\nabla}T = D_T {\bm\nabla}^2 T({\bm x},t)\ ,
\label{eq:2.35b}
\ee
\ese
with $D_T = \kappa/c_p$ the heat diffusivity; see Appendix~\ref{app:C} for a derivation. We note that the heat equation
(\ref{eq:2.35b}) can be written as an equation for the entropy per particle, i.e., the heat mode proper, by using
Eq.~(\ref{eq:A.6b}).

Equations~(\ref{eqs:2.35}), when supplemented by initial conditions $\delta T({\bm x},t=0)$ and ${\bm u}_{\perp}({\bm x},t=0)$,
describe the time evolution of macroscopic perturbations about the NESS characterized by ${\bm\nabla} T = \text{const.}$
and ${\bm u}_{\perp} = 0$. They remain valid if we replace the averaged quantities $T$ and ${\bm u}_{\perp}$ by their
operator-valued fluctuating counterparts ${\hat T}$ and ${\hat{\bm u}}_{\perp}$ that are moments of the operator-valued
$\mu$-space distribution ${\hat f}({\bm p},{\bm x},t)$ instead of its average $f_{\bm p}({\bm x},t)$, provided one adds
appropriate fluctuating, or Langevin, forces.\cite{Landau_Lifshitz_VI_1987} The linearized fluctuating quantum 
Navier-Stokes equations in the absence of a temperature gradient were derived in Ref.~\onlinecite{Kirkpatrick_Belitz_2022}
from a linearized quantum kinetic equation by means of projector techniques. The above discussion provides the
desired generalization to a NESS. Performing Fourier transforms in space and time, and choosing the coordinate
system as in Appendix~\ref{app:A} (see Fig.~\ref{fig:3}), the equations read
\bse
\label{eqs:2.36}
\be
\left(-i\omega + \nu {\bm k}^2\right) {\hat{u}}_{\perp}({\bm k},\omega) = {\hat P}_{\perp}({\bm k},\omega)\ ,\qquad\quad
%\nonumber\\
\label{eq:2.36a}
\ee
\be
\left(-i\omega + D_T {\bm k}^2 \right) {\hat T}({\bm k},\omega) + \perpgradT {\hat{u}}_{\perp}({\bm k},\omega) =  
{\hat Q}({\bm k},\omega)\ .
\label{eq:2.36b}
\ee
\ese
Here and in what follows we write $\hat{\bm k}_{\perp} \equiv \hat{\bm k}_{\perp}^{(1)}$, with
$\hat{\bm k}_{\perp}^{(1)}$ as defined in Appendix~\ref{app:A}, and $u_{\perp} \equiv {\bm u}_{\perp}\cdot{\hat{\bm k}}_{\perp}$. 
The fluctuating force operators
${\hat P}_{\perp}$ and ${\hat Q}$ have zero mean and are assumed to be Gaussian distributed. The second
moments of the distributions can be determined from the correlations of the more general $\mu$-space Langevin
operator that were determined in Ref.~\onlinecite{Kirkpatrick_Belitz_2022}. ${\hat Q}$ is related to the fluctuating
heat current ${\hat{\bm q}}_{\text{L}}$ defined in Ref.~\onlinecite{Kirkpatrick_Belitz_2022} by
\bse
\label{eqs:2.37}
\be 
{\hat Q}({\bm k},\omega) =  - i{\bm k}\cdot{\hat{\bm q}}_{\text{L}}({\bm k},\omega)/c_p\ , 
\label{eq:2.37a}
\ee
Similarly, the fluctuating force operator ${\hat P}_{\perp}$ is related to the fluctuating stress tensor ${\hat\tau}_{\text{L}}$
in Ref.~\onlinecite{Kirkpatrick_Belitz_2022} by 
\be
{\hat P}_{\perp}({\bm k},\omega) = \frac{-i}{\rho}\,{\hat k}_{\perp}^i k^j ({\hat\tau}_{\text{L}})_{ij}({\bm k},\omega) 
\label{eq:2.37b}
\ee
\ese
The anticommutator ($[\ ,\ ]_+$) and commutator ($[\ ,\ ]_-$) correlations, respectively, of $\hat Q$ are obtained from
Eq.~(3.24b) in that reference, and those of $\hat P_{\perp}$ from Eq.~(3.24a). We find\cite{hbar_footnote}
\begin{widetext}
\bse
\label{eqs:2.38}
\bea
\frac{1}{2}\left\langle\left[{\hat Q}({\bm k}_1,\omega_1),{\hat Q}({\bm k}_2,\omega_2)\right]_{\pm}\right\rangle &=& 
     2\pi \delta(\omega_1 + \omega_2)\,V\delta_{{\bm k}_1,-{\bm k}_2}\,\frac{D_T}{c_p}\,{\bm k}_1^2 \, \omega_1 T 
     c_{\pm}(\omega_1/2T)\ ,
\label{eq:2.38a}\\
\frac{1}{2}\left\langle\left[{\hat P}_{\perp}({\bm k}_1,\omega_1),{\hat P}_{\perp}({\bm k}_2,\omega_2)\right]_{\pm}\right\rangle &=& 
     2\pi \delta(\omega_1 + \omega_2)\,V\delta_{{\bm k}_1,-{\bm k}_2}\,\frac{\nu}{\rho}\,k_1^2 \, \omega_1  
   c_{\pm}(\omega_1/2T)   \ ,
\label{eq:2.38b}
\eea
where $T$ is the spatially averaged temperature and
\be
c_{\pm}(\Omega) =  \begin{cases}       \coth\Omega                          & \text{for} \quad +  \\
                                                                1                                          & \text{for} \quad -                                   
     \end{cases} 
     \label{eq:2.38c}
\ee
\ese
\end{widetext} 
The cross correlations vanish,
\be
\left\langle\left[{\hat Q}({\bm k}_1,\omega_1), {\hat P}_{\perp}({\bm k}_2,\omega_2)\right]_{\pm}\right\rangle = 0\ .
\label{eq:2.39}
\ee
Here we assume that the fluctuating force correlations in a NESS have the same form as in equilibrium. For arguments supporting
this assumption see, e.g., Refs.~\onlinecite{Ronis_Procaccia_Machta_1980} and \onlinecite{Kirkpatrick_Dorfman_2015}, 
and the discussion in Sec.~\ref{sec:V}.

\subsection{Kinetic equation for the collisionless regime}
\label{subsec:II.D}

As written, with frequency and wave-number independent transport coefficients $\DT$ and $\nu$, the
linearized fluctuating quantum Navier-Stokes equations (\ref{eqs:2.36}), together with Eqs.~(\ref{eqs:2.37}) - (\ref{eq:2.39}),
are valid in the hydrodynamic regime $\vF k < 1/\tau$, with $\tau$ the relevant relaxation time, that is dominated by collisions
between the quasiparticles, as is made explicit by the Chapman-Enskog expansion. (See, however, Appendix~\ref{app:D}). 
Since $\tau$ diverges as $T\to 0$, the hydrodynamic regime shrinks with decreasing temperature. In the collisionless regime 
in the opposite limit, $\vF k > 1/\tau$, which governs the asymptotic low-temperature behavior, 
we need to go back to the Uehling-Uhlenbeck equation (\ref{eq:2.5}) with the right-hand side replaced
by zero. The corresponding equation for the operator-valued distribution ${\hat f}$ is
\be
\partial_t {\hat f}({\bm p},{\bm x},t) + {\bm v}_{\bm p}\cdot{\bm\nabla}_{\bm x}{\hat f}({\bm p},{\bm x},t) = {\hat{\tilde F}}_{\text{L}}({\bm p},{\bm x},t)\ .
\label{eq:2.40}
\ee
Here ${\hat{\tilde F}}_{\text{L}}$ is an operator-valued Langevin force that is related to the fluctuating force ${\hat F}_{\text{L}}$ from
Eq.~(2.6a) in Ref.~\onlinecite{Kirkpatrick_Belitz_2022} by
\be
{\hat{\tilde F}}_{\text{L}}({\bm p},{\bm x},t) = w({\bm p})\,{\hat F}_{\text{L}}({\bm p},{\bm x},t)
\label{eq:2.41}
\ee
with
\be
w({\bm p}) = -\partial f_{\bm p}^{\text{eq}}/\partial\epsilon_p = \frac{1}{4T\cosh^2(\xi_p/2T)}
\label{eq:2.42}
\ee
with  $\xi_p = \epsilon_p - \mu$. ${\hat F}_{\text{L}}$ is Gaussian distributed with zero mean; the second moment of its 
distribution was determined in Ref.~\onlinecite{Kirkpatrick_Belitz_2022} and is given again in Eqs.~(\ref{eqs:2.48}) below.

Now consider the local equilibrium distribution from Eq.~(\ref{eq:2.7a}) and write
\be
{\hat f}({\bm p},{\bm x},t) = f_{\bm p}^{(0)}({\bm x},t) + \delta {\hat f}({\bm p},{\bm x},t)\ .
\label{eq:2.43}
\ee
We next anticipate that (1) the full distribution function depends on ${\bm p}$ only via the combination
${\bm c}({\bm x},t) = {\bm v}_{\bm p} - {\bm u}({\bm x},t)$ (see Eq.~(\ref{eq:2.26a})), and (2) we will eventually
sum over ${\bm p}$ in order to calculate observables. This suggests writing the streaming term in Eq.~(\ref{eq:2.40}) as
\bea
{\bm v}_{\bm p}\cdot{\bm\nabla}_{\bm x}{\hat f}({\bm p},{\bm x},t) &=& 
              {\bm c}({\bm x},t)\cdot{\bm\nabla}_{\bm x} {\hat f}(m({\bm c}+{\bm u}),{\bm x},t)
               \nonumber\\
&&  \hskip -20pt   + {\bm u}({\bm x},t)\cdot{\bm\nabla}_{\bm x} {\hat f}(m({\bm c}+{\bm u}),{\bm x},t)\ .\qquad
\label{eq:2.44}
\eea
The ${\bm u}\cdot{\bm\nabla}$ term is already linear in the fluctuations, so to linear order we can replace
${\hat f}$ in that term by the local equilibrium distribution with ${\bm u}=0$, the fluctuating chemical potential replaced
by its average value $\mu$, and $T({\bm x},t)$ replaced by the externally imposed linear temperature profile. 
If we again neglect pressure fluctuations\cite{zero-sound_footnote} the ${\bm\nabla}_{\bm x}{\hat f}$ term
evaluated at constant pressure becomes
%
% See p. 155-17 ff
\be
{\bm\nabla}_{\bm x}\,\frac{1}{e^{\xi_p/T({\bm x})} + 1} = w({\bm p})\,a_s({\bm p})\,\frac{1}{T}\,{\bm\nabla}T\ ,
\label{eq:2.45}
\ee
with $a_s({\bm p}) = \xi_{\bm p} - sT/n$ from Eq.~(\ref{eq:A.1f}).
We see that the kinetic equation contains the same ${\bm u}\cdot{\bm\nabla}T$ term as the Navier-Stokes equations,
and we again keep only the coupling to the transverse velocity fluctuations. That is, we ignore all other effects of the 
temperature gradient and write ${\hat f}({\bm p},{\bm x},t) = f_{\bm p}^{\text{eq}} + \delta {\hat f}({\bm p},{\bm x},t)$ in all
other terms in the kinetic equation. Defining a function ${\hat\phi}$ by
\be
\delta{\hat f}({\bm p},{\bm x},t) = w({\bm p})\,{\hat\phi}({\bm p},{\bm x},t)\ ,
\label{eq:2.46}
\ee
and linearizing the kinetic equation, we find a linearized version of Eq.~(\ref{eq:2.40})
appropriate for a fluid in a NESS characterized by a constant temperature gradient:
\bea
\left(\partial_t + {\bm v}_{\bm p}\cdot{\bm\nabla}_{\bm x}\right) {\hat\phi}({\bm p},{\bm x},t) &=& {\hat F}_{\text{L}}({\bm p},{\bm x},t)
\nonumber\\
&& \hskip -30pt   - {\hat u}_{\perp}({\bm x},t)\,\frac{\perpgradT}{T}\,a_s({\bm p})\ .\qquad
\label{eq:2.47}
\eea

Equation~(\ref{eq:2.47}) generalizes Eq.~(2.6b) in Ref.~\onlinecite{Kirkpatrick_Belitz_2022} 
to the case of a constant temperature gradient while neglecting the LFL interaction. 
The correlations of the Langevin force ${\hat F}_{\text{L}}$ were given in Sec.~II.C of Ref.~\onlinecite{Kirkpatrick_Belitz_2022}
and we list them here again for completeness:
\bse
\label{eqs:2.48}
\bea
\frac{1}{2} \Big\langle \left[{\hat F}_{\text{L}}({\bm p}_1,{\bm k}_1,\omega_1), {\hat F}_{\text{L}}({\bm p}_2,{\bm k}_2,\omega_2)\right]_{\pm}\Big\rangle  
               &=& 2\pi \delta(\omega_1+\omega_2)\,
\nonumber\\
&&\hskip -120pt  \times V \delta_{{\bm k}_1+{\bm k}_2,0}\,\Psi_{\pm}({\bm p}_1,{\bm p}_2; {\bm k}_1,\omega_1)\ ,
\label{eq:2.48a}
\eea   
where
\bea
\Psi_{\pm}({\bm p}_1,{\bm p}_2; {\bm k},\omega) &=& \frac{-\omega}{2T}\,c_{\pm}(\omega/2T) \left[\Lambda({\bm p}_1) + \Lambda({\bm p}_2)\right]\,
\nonumber\\
&&  \times V \delta_{{\bm p}_1,{\bm p}_2}\,\frac{T}{w({\bm p}_1)} \ ,
%      \nonumber\\
\label{eq:2.48b}
%\Psi_{+}({\bm p}_1,{\bm p}_2; {\bm k},\omega) &=& \coth\left(\frac{\omega}{2T}\right) \Psi_{-}({\bm p}_1,{\bm p}_2; {\bm k},\omega)\ .
%\label{eq:2.48c}
\eea
\ese          
with $c_{\pm}$ from Eq.~(\ref{eq:2.38c}). Here $\Lambda({\bm p})$ is the same linearized collision operator as in Eq.~(\ref{eq:2.25}).

\section{A Fermi Liquid in a NESS I: Correlation functions}
\label{sec:III}

We now are in a position to calculate the temperature correlation functions for a quantum fluid subject to a
constant temperature gradient, i.e., the quantum counterparts of Eq.~(\ref{eq:1.2}). As mentioned in the
Introduction, we need to distinguish between anticommutator ($[\  ,\ ]_+$), or symmetric, correlation functions 
$S^{\text{sym}}$ that are also referred to as fluctuation functions, and 
commutator ($[\ ,\ ]_-$), or antisymmetric, correlation functions $\chi''$; see Appendix~\ref{app:B} for a summary of definitions. 
They are defined by
\begin{widetext}
\bse
\label{eqs:3.1}
\bea
\frac{1}{2}\,\left\langle\left[\delta {\hat A}({\bm k}_1,\omega_1), \delta {\hat B}({\bm k}_2,\omega_2)\right]_{+}\right\rangle
&=& V\delta_{{\bm k}_1,-{\bm k}_2}\,2\pi \delta(\omega_1+\omega_2)\,\Ssym_{AB}({\bm k}_1,\omega_1)
\label{eq:3.1a}\\
\frac{1}{2\hbar}\,\left\langle\left[\delta {\hat A}({\bm k}_1,\omega_1), \delta {\hat B}({\bm k}_2,\omega_2)\right]_{-}\right\rangle
&=& V\delta_{{\bm k}_1,-{\bm k}_2}\,2\pi \delta(\omega_1+\omega_2)\, \chi_{AB}''({\bm k}_1,\omega_1) 
\label{eq:3.1b}
\eea
\ese
\end{widetext}
where the observables ${\hat A}$ and ${\hat B}$ can stand for either ${\hat T}$ or ${\hat u}_{\perp}$. 
For the purpose of Eq.~(\ref{eq:3.1b}) we have restored $\hbar$ (see Ref.~\onlinecite{hbar_footnote}).
Within the approximations that we are employing throughout this paper,\cite{approximations_footnote}
$\Ssym$ and $\chi''$ are related by the  factor from Eq.~(\ref{eq:2.38c}):
\be
\Ssym_{AB}({\bm k},\omega) = \chi''_{AB}({\bm k},\omega)\,\coth(\omega/2T)\ .
\label{eq:3.2}
\ee
The symmetrized correlation functions $\Ssym_{AB}$ are observable by means of scattering experiments.\cite{Forster_1975}
The physical meaning of the antisymmetrized correlation functions $\chi''_{AB}$ is {\em a priori} less obvious.
In an equilibrium system, where the correlations are generically short-ranged, they describe the linear response of the system to external
fields. That is, the equilibrium fluctuations determine the linear response, which to second order in the external
field yields the energy dissipated by the system. This is the content of the fluctuation-dissipation 
theorem.\cite{Nyquist_1928, Callen_Welton_1951} In a NESS, the relation (\ref{eq:3.2}) 
between commutator and anticommutator correlations still holds, but the
commutator correlations functions no longer describe the linear response and the usual fluctuation-dissipation
theorem breaks down. We will discuss the physical meaning of the commutator correlation functions in Sec.~\ref{subsec:IV.B}.

In addition, we need to distinguish between the hydrodynamic regime $\omega\tau < 1$ (or, equivalently, $\vF k\tau<1$), 
where the Chapman-Enskog derivation of the quantum Navier-Stokes equations is valid, and the collisionless regime 
$\omega\tau > 1$, where one has to work with the $\mu$-space 
kinetic equation.\cite{Navier-Stokes_validity_footnote} Figure~\ref{fig:2} illustrates the various frequency/energy regimes.
In an ordinary Fermi liquid the relaxation rate is $1/\tau \sim T^2/\epsilonF$,\cite{Landau_Lifshitz_IX_1991}
with $\epsilon_F$ the Fermi energy.
In more exotic Fermi systems the rate can scale as a smaller power of $T$, but there are good arguments for
$T$ being an upper bound on $1/\tau$,\cite{Hartnoll_Mackenzie_2022} so $1/\tau \alt T$ always. The collisionless
regime thus is divided into two subregimes where $\omega<T$ and $\omega > T$, respectively. The frequency
scales quadratically with the wave number $k$ in the hydrodynamic regime, and linearly in the collisionless regime.
\begin{figure}[h]
\includegraphics[width=8.5cm]{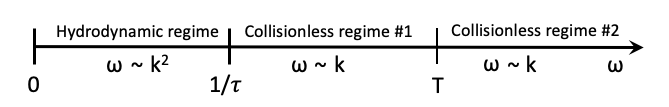}
\caption{Relevant frequency/energy regimes.}
\label{fig:2}
\end{figure}

\subsection{Correlation functions in the hydrodynamic regime}
\label{subsec:III.A}

To find the desired correlations in the hydrodynamic regime we solve Eqs.~(\ref{eqs:2.36}) for $\delta{\hat T}$
and ${\hat u}_{\perp}$ in terms of the fluctuating forces. This yields
\begin{widetext}
\bse
\label{eqs:3.3}
\bea
{\hat u}_{\perp}({\bm k},\omega) &=&\frac{1}{-i\omega + \nu k^2}\,{\hat P}_{\perp}({\bm k},\omega)\ ,
\label{eq:3.3a}\\
\delta{\hat T}({\bm k},\omega) &=& \frac{1}{-i\omega + D_{\text{T}}k^2}\left[{\hat Q}({\bm k},\omega) - \frac{\perpgradT}{-i\omega + \nu k^2}\,{\hat P}_{\perp}({\bm k},\omega)\right]
\nonumber\\
&=&  \frac{1}{-i\omega + D_{\text{T}}k^2}\,{\hat Q}({\bm k},\omega) - \frac{\perpgradT}{-i\omega + D_{\text{T}}k^2}\,{\hat u}_{\perp}({\bm k},\omega)\ .
\label{eq:3.3b}
\eea
\ese
We see that the equilibrium contribution to the temperature fluctuations is given by the fluctuating force $\hat Q$, whereas the
non-equilibrium contribution is given in terms of the shear-velocity fluctuations. The correlation functions of the latter are given 
by their equilibrium expressions,\cite{local_equilibrium_footnote} viz.,
\bse
\label{eqs:3.4}
\bea
\Ssym_{u_{\perp}u_{\perp}}({\bm k},\omega) &=& \frac{2T}{\rho}\,\frac{\nu k^2}{\omega^2 + (\nu k^2)^2}\ ,
\label{eq:3.4a}\\
\Ssym_{u_{\perp}u_{\perp}}({\bm k}) &=&    \int_{-\infty}^{\infty} \frac{d\omega}{2\pi}\,\Ssym_{u_{\perp}u_{\perp}}({\bm k},\omega)  =  T/\rho\ ,
\label{eq:3.4b}\\
\chi_{u_{\perp}u_{\perp}}''({\bm k},\omega) &=& \frac{1}{\rho}\,\frac{\omega \nu k^2}{\omega^2 + (\nu k^2)^2}\ ,
\label{eq:3.4c}\\
\chi_{u_{\perp}u_{\perp}}({\bm k}) &=&  \int_{-\infty}^{\infty} \frac{d\omega}{\pi}\,\chi_{u_{\perp}u_{\perp}}''({\bm k},\omega))/\omega            = 1/\rho\ .
\label{eq:3.4d}
\eea
\ese
Here we have used Eq.~(\ref{eq:2.38b}) as well as the fact that in the hydrodynamic regime the $\coth$ in Eq.~(\ref{eq:3.2}) is 
effectively $\coth(\omega/2T) \approx 2T/\omega$. Using Eq.~(\ref{eq:2.38a}) we find for the temperature correlation functions
\bse
\label{eqs:3.5}
\bea
\Ssym_{TT}({\bm k},\omega) &=& \frac{2T}{\omega^2 + D_{\text{T}}^2 k^4} \biggl[\frac{T}{c_p}\,D_{\text{T}}k^2 
     + \frac{\perpgradT^2}{\rho}\, \frac{\nu k^2}{\omega^2 + (\nu k^2)^2} \biggr]\ ,     
\label{eq:3.5a}\\
\chi_{TT}''({\bm k},\omega) &=& \frac{\omega}{\omega^2 + D_{\text{T}}^2 k^4} \biggl[\frac{T}{c_p}\,D_{\text{T}}k^2 
     + \frac{\perpgradT^2}{\rho}\,\frac{\nu k^2}{\omega^2 + (\nu k^2)^2} \biggr]\ , \qquad\quad
\label{eq:3.5b}
\eea
\ese
Equations~(\ref{eqs:3.5}) demonstrate the $\omega \sim k^2$ scaling that is characteristic of the hydrodynamic regime.
The corresponding static correlation functions are
\bse
\label{eqs:3.6}
\bea
\Ssym_{TT}({\bm k}) &=&  \int_{-\infty}^{\infty} \frac{d\omega}{2\pi}\,\Ssym_{TT}({\bm k},\omega)
                                         = { \frac{T^2}{c_p} + \frac{T \perpgradT^2}{\rho D_{\text{T}} (\nu + D_{\text{T}}) \, k^4}\ ,}
\label{eq:3.6a}\\
\chi_{TT}({\bm k}) &=& \int_{-\infty}^{\infty} \frac{d\omega}{\pi}\,\chi_{TT}''({\bm k},\omega)/\omega
                                        = { \frac{T}{c_p} + \frac{\perpgradT^2}{\rho D_{\text{T}} (\nu + D_{\text{T}}) \, k^4}\ .}
\label{eq:3.6b}
\eea
\ese
\end{widetext}
This is the same result as in the classical case,\cite{Kirkpatrick_Cohen_Dorfman_1982c, Dorfman_Kirkpatrick_Sengers_1994, Ortiz_Sengers_2007}
where the symmetrized correlation function $\Ssym_{TT}$ becomes identical with the van Hove function $S_{TT}$.\cite{Forster_1975}
This was to be expected since the structure of the quantum Navier-Stokes equations is the same as that of the classical ones.
Note that the equilibrium contribution to $\Ssym_{TT}$ is $T^2/c_p$, rather than $T^2/c_V$ (see Eqs.~(\ref{eqs:1.1})), 
since we have neglected the pressure fluctuations. If one keeps the latter, then the Brioullin, or sound-wave, peaks in
the structure factor contribute to the sum rule and change $c_p$ to $c_V$, just as in the classical case.\cite{Forster_1975}

For the mixed correlation functions we obtain\cite{kperp_footnote}
\bse
\label{eqs:3.7}
\bea
\Ssym_{u_{\perp}T}({\bm k},\omega) &=& - \perpgradT\,\frac{1}{\rho}\, \frac{\nu k^2}{\omega^2 + (\nu k^2)^2}\,\frac{2T}{i\omega + \DT k^2}\ ,
\nonumber\\
\label{eq:3.7a}\\
\Ssym_{T u_{\perp}}({\bm k},\omega) &=& \Ssym_{u_{\perp}T}(-{\bm k},-\omega)\ ,
\label{eq:3.7b}\\
\Ssym_{u_{\perp}T}({\bm k}) &=& \Ssym_{T u_{\perp}}(-{\bm k}) = - \perpgradT\, \frac{T}{\rho}\,\frac{1}{(\nu + \DT)k^2}\ ,
\nonumber\\
\label{eq:3.7c}\\
\chi_{u_{\perp}T}''({\bm k},\omega) &=& - \perpgradT\,\frac{1}{\rho}\, \frac{\omega\nu k^2}{\omega^2 + (\nu k^2)^2}\,\frac{1}{i\omega + \DT k^2}\ ,
\nonumber\\
\label{eq:3.7d}\\
\chi_{Tu_{\perp}}''({\bm k},\omega) &=& - \chi_{u_{\perp}T}''(-{\bm k},-\omega)
\label{eq:3.7e}\\
\chi_{u_{\perp}T}({\bm k}) &=& \chi_{Tu_{\perp}}(-{\bm k}) = - \perpgradT\, \frac{1}{\rho}\,\frac{1}{(\nu + \DT)k^2}\ .
\nonumber\\
\label{eq:3.7f}
\eea
\ese
Equations~(\ref{eq:3.7b}) and (\ref{eq:3.7e}) reflect a general symmetry property of anticommutator and commutator correlation functions,
respectively.\cite{Kadanoff_Martin_1963}

In all cases the static correlation functions are related by
\bse
\label{eqs:3.8}
\be
S_{AB}^{\text{sym}}({\bm k}) = T\,\chi_{AB}({\bm k})\,\qquad (A,B = T, u_{\perp})\ ,
\label{eq:3.8a}
\ee
\bigskip
as must be the case given Eq.~(\ref{eq:3.2}). For later reference we write the left-hand side more explicitly, using Eq.~(\ref{eq:3.1a}),
\be
\frac{1}{2V} \int \frac{d\omega}{2\pi}\,\frac{d\omega'}{2\pi}\,\langle [\delta{\hat A}({\bm k},\omega) , \delta{\hat B}(-{\bm k},\omega')]_+ \rangle = T\,\chi_{AB}({\bm k})\ .
\label{eq:3.8b}
\ee
\ese

One important consequence of the structure of Eqs.~(\ref{eqs:3.3}) - (\ref{eqs:3.7}) is the following: Since the $u_{\perp}$-correlation
functions are the same as in (local) equilibrium, see Eqs.~(\ref{eqs:3.4}), $\chi_{u_{\perp}u_{\perp}}''$ determines the linear response
of the system to a field conjugate to $u_{\perp}$. Since the shear fluctuations completely determine the non-equilibrium part of the
temperature fluctuations, this implies that all of the non-equilibrium effects expressed in Eqs.~(\ref{eqs:3.6}) - (\ref{eqs:3.7}) can be
probed via the linear response to an initial shear perturbation. We will elaborate on this observation in Sec.~\ref{sec:IV}.

%\smallskip
\subsection{Correlation functions in the collisionless regime}
\label{subsec:III.B}

\subsubsection{Approximate solution of the kinetic equation}
\label{subsubsec:III.B.1}

We determine the behavior in the collisionless regime by solving the linearized kinetic equation (\ref{eq:2.47}). A Fourier transform in
space and time yields {(see Eq.~(\ref{eq:2.46}))}
\begin{widetext}
\bse
\label{eqs:3.9}
\bea
\delta{\hat f}({\bm p},{\bm k},\omega) &=& w({\bm p})\,{\hat\phi}({\bm p},{\bm k},\omega) 
\nonumber\\
&=& w({\bm p})\,G_0({\bm p},{\bm k},\omega) \biggl[{\hat F}_{\text{L}}({\bm p},{\bm k},\omega)  - \hat{u}_{\perp}({\bm k},\omega)\, \frac{\perpgradT}{T}\,a_s({\bm p})\biggr]
\label{eq:3.9a}
\eea
with
\be
G_0({\bm p},{\bm k},\omega) = \frac{i}{\omega - {\bm k}\cdot{\bm p}/m + i0}
\label{eq:3.9b}
\ee
\ese
a Green function. Here $i0$ indicates a positive infinitesimal imaginary quantity. The temperature fluctuations are
given by Eq.~(\ref{eq:A.4d}). Substituting the solution (\ref{eq:3.9a}) of the kinetic equation, we have
\bse
\label{eqs:3.10}
\be
\delta{\hat T}({\bm k},\omega) = \frac{1}{c_V}\,\frac{1}{V}\sum_{\bm p} w({\bm p})\,a_5({\bm p})\,G_0({\bm p},{\bm k},\omega)\,{\hat F}_{\text{L}}({\bm p},{\bm k},\omega)
      - \perpgradT\,{\hat u}_{\perp}({\bm k},\omega)\,\tau({\bm k},\omega) \ ,
\label{eq:3.10a}
\ee
where
\be
\tau({\bm k},\omega) = \frac{1}{c_V T}\,\frac{1}{V}\sum_{\bm p} w({\bm p})\,a_5({\bm p})\,a_s({\bm p})\,G_0({\bm p},{\bm k},\omega)\hskip 30pt
\label{eq:3.10b}
\ee
In the low-temperature limit this becomes
\be
\tau({\bm k},\omega) = \frac{\NF T}{c_V}\,\frac{\pi^2}{6}\,\frac{-i}{\vF k}\,\log\left(\frac{1 - \omega/\vF k - i0}{-1 - \omega/\vF k - i0}\right) 
+ O(T^3)\ ,
\label{eq:3.10c}
\ee
\ese
\end{widetext}
with $\NF$ the density of states at the Fermi surface.
The functions $a_5$ and $a_s$ are defined in Eqs.~(\ref{eqs:A.1}), and in Eq.~(\ref{eq:3.10c}) we have evaluated
the integral to leading order for $T\to 0$. 
Again the equilibrium part of the $\delta T$-correlation function is given by the correlation of the 
fluctuating force, whereas the non-equilibrium part is given by the correlation of the transverse velocity;
see Eq.~(\ref{eq:3.3b}) for the analogous structure in the hydrodynamic regime.
To calculate the latter we combine Eqs.~(\ref{eq:A.4c}), (\ref{eq:2.46}), (\ref{eq:3.9a}), and (\ref{eqs:2.48}). 
To leading order as $T\to 0$ we find
\begin{widetext}
\be
\frac{1}{2} \Big\langle \left[{\hat u}_{\perp}({\bm k}_1,\omega_1), {\hat u}_{\perp}({\bm k}_2,\omega_2)\right]_{\pm}\Big\rangle  
               = 2\pi \delta(\omega_1+\omega_2)\,
V\,\delta_{{\bm k}_1,-{\bm k}_2}\,\frac{\pi}{\rho^2}\,\omega_1\,\frac{1}{V}\sum_{\bm p} w({\bm p})\,
     \left(\hat{\bm k}_{1\perp}\cdot{\bm p}\right)^2 \delta(\omega_1 - {\bm k}_1\cdot{\bm p}/m)\,c_{\pm}(\omega_1/2T)\ .
\label{eq:3.11}
\ee
Here we have used the low-temperature limiting procedure given in Eqs.~(3.7) of Ref.~\onlinecite{Kirkpatrick_Belitz_2022}.
For the temperature-temperature correlations this yields
\bse
\label{eqs:3.12}
\bea
\chi''_{TT}({\bm k},\omega) &=& \frac{\pi}{c_V^2}\,\frac{1}{V} \sum_{\bm p} w({\bm p})\,\left(a_5({\bm p})\right)^2 \omega\,\delta(\omega - {\bm k}\cdot{\bm p}/m)
\nonumber\\
&& +\, \omega\,\frac{\pi}{4}\,\frac{(\perpgradT)^2}{\vF^2 k^2}\,\frac{\kF^2}{\rho^2}\,\frac{1}{V}\sum_{\bm p} w({\bm p}) \left(1 - (\hat{\bm k}\cdot\hat{\bm p})^2\right) 
     \left[\log^2 \left\vert\frac{1 - \hat{\bm k}\cdot\hat{\bm p}}{1 + \hat{\bm k}\cdot\hat{\bm p}}\right\vert + \pi^2\right]\,\delta(\omega - {\bm k}\cdot{\bm p}/m) \qquad
\label{eq:3.12a}
\eea
and
\be
S^{\text{sym}}_{TT}({\bm k},\omega) = \chi''_{TT}({\bm k},\omega)\, \coth(\omega/2T)\ .
\label{eq:3.12b}
\ee
\ese
\end{widetext}

\subsubsection{Static correlation functions}
\label{subsubsec:III.B.2}

We can now determine the desired static correlation functions. From (\ref{eq:3.12a}) we find, using the first equality in Eq.~(\ref{eq:3.6b}),
and after some algebra,
\be
\chi_{TT}({\bm k}) = \frac{1}{\NF}\,\frac{3}{\pi^2} \left[1 + \frac{\pi^2}{12}\,(2\pi^2-3)\,\frac{\perpgradT^2}{\epsilonF^2 k^2}\right]\ .
\label{eq:3.13}
\ee
Here we have used the fact that in the low-temperature limit the specific heat is $c_p \approx c_V \approx (\pi^2/3)\,\NF T$.

The non-equilibrium contribution has a weaker singularity than in the hydrodynamic regime ($1/k^2$ rather than $1/k^4$)
since the transverse velocity modes are now ballistic rather than diffusive.

For the symmetrized correlation function the additional factor of $\coth(\omega/2T)$ forces us
to distinguish between two subregimes:

\paragraph{$1/\tau < \vF k < T$} This regime exists provided $1/\tau \ll T$. Here  $\coth(\omega/2T) \approx 2T/\omega$, and
we obtain the same relation between $\Ssym$ and $\chi$ as in the hydrodynamic regime, 
\bse
\label{eqs:3.14}
\bea
\Ssym_{TT}({\bm k}) &=& T\,\chi_{TT}({\bm k})
\nonumber\\
&=& \frac{T}{\NF}\,\frac{3}{\pi^2} \left[1 + s^{(a)}\,\frac{\perpgradT^2}{\epsilonF^2 k^2}\right]\ .\qquad
\label{eq:3.14a}
\eea
with
\be
s^{(a)} = \frac{\pi^2}{12}\,(2\pi^2-3)\ .
\label{eq:3.14b}
\ee
\ese

\paragraph{$T < \vF k$} Here $\coth(\omega/2T) \approx \sgn\omega$, and we find
\bse
\label{eqs:3.15}
\be
\Ssym_{TT}({\bm k}) = \frac{\vF k}{\NF}\,\frac{3}{\pi^2} \biggl[1 +
s^{(b)}\,\frac{\perpgradT^2}{\epsilonF^2 k^2}\biggr]\ .
\label{eq:3.15a}
\ee
where
\be
s^{(b)} = \frac{\pi^2}{128}\,(3\pi^2 + 16 \log 2 - 4)\ .
\label{eq:3.15b}
\ee
\ese
We see that in this asymptotic low-temperature regime the relation between $\Ssym$ and $\chi$ is, apart from
factors of $O(1)$, $\Ssym_{TT}({\bm k}) \approx \vF k\,\chi_{TT}({\bm k})$. Consequently, the non-equilibrium
contribution to $\Ssym_{TT}$ diverges as $1/k$.

As mentioned in Sec.~\ref{subsec:II.C}, these results reflect the coupling of the temperature fluctuations to the 
transverse velocity fluctuations only. All other soft modes in the low-temperature limit are also ballistic and hence
cannot change the leading scaling behavior, but they change the prefactor of the $k\to 0$ singularity.
How many soft modes exist at zero temperature depends on the values of the Landau Fermi-liquid
parameters, see the discussion in Ref.~\onlinecite{Belitz_Kirkpatrick_2022}. In particular we note that
the longitudinal velocity fluctuations are soft in a neutral Fermi liquid (where they constitute longitudinal
zero sound), but not in a charged Fermi liquid, where they turn into the massive plasmon. Our procedure,
which ignores pressure fluctuations, is thus better justified for conduction electrons in a metal than for
a neutral Fermi liquid.

In Appendix~\ref{app:D} we show that the results obtained from the Navier-Stokes equations are
consistent with Eqs.~(\ref{eq:3.13}) and (\ref{eqs:3.15}) if one uses the fact that the diffusion coefficients
$\DT$ and $\nu$ effective scale as $1/k$ in the collisionless regime. We note, however, that in the
hydrodynamic regime the Navier-Stokes equations capture all of the relevant soft modes, while in
the collisionless regime they do not (and neither does our approximate solution of the kinetic equation).

Static correlation functions that diverge in the limit of zero wave number are familiar from systems with
a spontaneously broken continuous symmetry, where they represent Goldstone modes.\cite{Forster_1975}
Here, they appear as a property of the NESS, in the absence of any symmetry breaking. In either case
they signal a generalized rigidity of the system that results in long-ranged spatial correlations.\cite{Anderson_1984}
In the next section we discuss consequences of this generalized rigidity,
namely, the response to an initial applied shear and the anomalous propagation of a localized temperature perturbation.

\section{A Fermi Liquid in a NESS II: Response functions}
\label{sec:IV}

The symmetrized correlation functions discussed in Sec.~\ref{sec:III} are measurable by light scattering.
This is how the classical prediction expressed in Eq.~(\ref{eq:1.2}) was confirmed experimentally,
see, e.g., Ref.~\onlinecite{Sengers_Ortiz_Kirkpatrick_2016} and references therein. These are
difficult experiments even in classical fluids because of the very small scattering angles required.
As the temperature is lowered, the fluctuation effects become weaker, which makes the experimental
task even more difficult. It therefore is desirable to consider the effects of the long-range correlations
on the response of the system to external perturbations. In a NESS this is a more difficult task than in
equilibrium, since there is no simple fluctuation-dissipation theorem that relates the correlation
functions to the response functions. In this section we show that, remarkably, the commutator
correlation functions in a NESS are still related to response functions via the bilinear response
of products of observables to an external field conjugate to the transverse velocity. This shows
in particular that the long-range correlations discussed in Sec.~\ref{sec:III}, and the related 
generalized rigidity of the NESS, are encoded in the averaged hydrodynamic equations and are 
not related to fluctuating forces. 

We then discuss another manifestation of the generalized rigidity, namely, the anomalous 
propagation of a temperature pulse that results from an initial shear and temperature
perturbation. 

We finally discuss how response experiments allow for the measurement of the commutator
correlation functions via macroscopic driving terms that are experimentally controllable and
independent of the temperature.

\subsection{Linear response to an external shear velocity perturbation}
\label{subsec:IV.A}

In order to study the linear response of the system we consider the simplified Navier-Stokes
equations (\ref{eqs:2.36}) for the averaged quantities $u_{\perp} = \langle {\hat u}_{\perp}\rangle$
and $\delta T = \langle \delta{\hat T}\rangle$. The averaged fluctuating forces vanish, and we
add an external field $h_{u_{\perp}}$ conjugate to the shear velocity $u_{\perp}$. This amounts to simply 
shifting the transverse velocity by the field times the static $u_{\perp}$ susceptibility, which equals
$1/\rho$.\cite{TDGL_footnote} The equations then are
\bse
\label{eqs:4.1}
\be
\left(-i\omega + \nu k^2\right) u_{\perp}({\bm k},\omega) = \frac{\nu}{\rho}\, k^2 h_{u_{\perp}}({\bm k},\omega)\ ,
\label{eq:4.1a}
\ee
\bea
\left(-i\omega + D_T k^2\right) \delta T({\bm k},\omega) + \perpgradT u_{\perp}({\bm k},\omega) &=& 
\nonumber\\
&& \hskip -120pt \frac{1}{\rho}\perpgradT h_{u_{\perp}}({\bm k},\omega)\  . \qquad
\label{eq:4.1b}
\eea
\ese
They are easily solved to obtain response functions $X_{Tu_{\perp}}$ and $X_{u_{\perp}u_{\perp}}$ defined by (see also Appendix~\ref{app:B})
\bse
\label{eqs:4.2}
\bea
u_{\perp}({\bm k},\omega) &=& X_{u_{\perp}u_{\perp}}({\bm k},\omega)\,h_{u_{\perp}}({\bm k},\omega) \ ,
\label{eq:4.2a}\\
\delta T({\bm k},\omega) &=& X_{Tu_{\perp}}({\bm k},\omega)\,h_{u_{\perp}}({\bm k},\omega) \ .
\label{eq:4.2b}
\eea
\ese
We find
\bse
\label{eqs:4.3}
\bea
X_{u_{\perp}u_{\perp}}({\bm k},\omega) &=& \frac{1}{\rho}\,\frac{\nu k^2}{-i\omega + \nu k^2}\ .
\label{eq:4.3a}\\
X_{Tu_{\perp}}({\bm k},\omega) &=& \frac{1}{\rho}\,\perpgradT\,\frac{1}{-i\omega + \DT k^2}\,\frac{-i\omega}{-i\omega + \nu k^2}\ .
\nonumber\\
\label{eq:4.3b}
\eea
\ese
In equilibrium the spectra, {or spectral densities}, of these response functions, 
$X''({\bm k},\omega) = \left[X({\bm k},\omega + i\epsilon) - X({\bm k},\omega - i\epsilon)\right]/2i$,
would be identical with the commutator correlation functions $\chi''_{Tu_{\perp}}$ (for $X_{Tu_{\perp}}$) and 
$\chi''_{u_{\perp}u_{\perp}}$ (for $X_{u_{\perp}u_{\perp}}$). 
For $X_{u_{\perp}u_{\perp}}$ this still holds in a NESS, as we see by comparing Eqs.~(\ref{eq:4.3a}) and (\ref{eq:3.4c}). We have
\bse
\label{eqs:4.4}
\bea
X''_{u_{\perp}u_{\perp}}({\bm k},\omega) &=& \chi''_{u_{\perp}u_{\perp}}({\bm k},\omega) = \frac{1}{\rho}\,\frac{\omega\nu k^2}{\omega^2 + \nu^2 k^4}\ ,
\nonumber\\
\label{eq:4.4a}\\
X''_{u_{\perp}u_{\perp}}({\bm k}) &=&  \chi''_{u_{\perp}u_{\perp}}({\bm k}) = 1/\rho\ .
\label{eq:4.4b}
\eea
\ese
However, the spectrum of $X_{Tu_{\perp}}$, 
\bse
\label{eqs:4.5}
\be
X_{Tu_{\perp}}''({\bm k},\omega) = -\perpgradT \omega\,\frac{\nu \DT k^2 - \omega^2}{(\omega^2 + \DT^2 k^4)(\omega^2 + \nu^2 k^4)}\ ,
\label{eq:4.5a}
\ee
is not identical with $\chi_{Tu_{\perp}}''$, although the two functions show the same scaling behavior. 
In particular, the static response function vanishes,
\be
X_{Tu_{\perp}}({\bm k}) = \int \frac{d\omega}{\pi}\,\frac{X_{Tu_{\perp}}''({\bm k},\omega)}{\omega} = 0\ ,
\label{eq:4.5b}
\ee
\ese
whereas $\chi_{Tu_{\perp}}({\bm k})$ is nonzero, see Eq.~(\ref{eq:3.7f}).

As we will see, it is also useful to define an observable
\be
{\tilde T}({\bm k},\omega) = T({\bm k},\omega) - \frac{1}{\rho}\,\perpgradT\,\frac{1}{-i\omega + \DT k^2}\,h_{u_{\perp}}({\bm k},\omega)
\label{eq:4.6}
\ee
that obeys the equation
\be
\left(-i\omega + D_T k^2\right) \delta {\tilde T}({\bm k},\omega) + \perpgradT u_{\perp}({\bm k},\omega) = 0\ .
\label{eq:4.7}
\ee
\smallskip\par\noindent
Comparing Eqs.~(\ref{eq:4.7}) and (\ref{eq:4.1b}) we see that this is the heat equation with a
streaming term that contains the absolute shear velocity, whereas the streaming term in the equation for $\delta T$ 
contains the shear velocity relative to the external field $h_{u_{\perp}}$. 
The response of ${\tilde T}$ to the external field $h_{u_{\perp}}$ is given by a response function
\be
X_{{\tilde T}u_{\perp}}({\bm k},\omega) = \frac{1}{\rho}\,\perpgradT\,\frac{-1}{-i\omega + \DT k^2}\,\frac{\nu k^2}{-i\omega + \nu k^2}\ .
\label{eq:4.8}
\ee
Finally, we define the shear velocity relative to the field $h_{u_{\perp}}$,
\be
{\tilde u}_{\perp}({\bm k},\omega) = u_{\perp}({\bm k},\omega)  - \frac{1}{\rho}\,h_{u_{\perp}}({\bm k},\omega)\ ,
\label{eq:4.9}
\ee
which obeys
\be
\left(-i\omega + \nu k^2\right) {\tilde u}_{\perp}({\bm k},\omega) = i\omega\,\frac{1}{\rho}\,h_{u_{\perp}}({\bm k},\omega)\ ,
\label{eq:4.10}
\ee
and whose response to the field is given by
\be
X_{{\tilde u}_{\perp}u_{\perp}}({\bm k},\omega) = \frac{1}{\rho}\,\frac{i\omega}{-i\omega + \nu k^2}\ .
\label{eq:4.11}
\ee

Note that the field $h_{u_{\perp}}$ can be experimentally realized by means of an imposed initial shear
velocity. Suppose the field is switched on adiabatically in the distant past and switched off discontinuously
at time $t=0$:
\bse
\label{eqs:4.12}
\be
h_{u_{\perp}}({\bm k},t) = h_{u_{\perp}}({\bm k})\,e^{\epsilon t}\,\Theta(-t)
\label{eq:4.12a}
\ee
with $\epsilon>0$ infinitesimal. Then the field produces a shear velocity at $t=0$ (see Eqs.~(\ref{eq:B.7}) and 
(\ref{eq:4.4b}))
\be
u_{\perp}({\bm k},t=0) = \frac{1}{\rho}\,h_{u_{\perp}}({\bm k})\ .
\label{eq:4.12b}
\ee
\ese

\subsection{A relation between response functions and correlation functions in a NESS}
\label{subsec:IV.B}

We now show that in a NESS there still is a relation between the response functions and the
antisymmetric, or commutator, correlation functions. {In this subsection we restrict ourselves to the hydrodynamic regime.}

Consider the product of a temperature fluctuation and a shear-velocity fluctuation, and its response
to the external field $h_{u_{\perp}}$. We have
\begin{widetext}
\bea
\delta T({\bm k},\omega)\,u_{\perp}(-{\bm k},-\omega) &=& X_{Tu_{\perp}}({\bm k},\omega)\,X_{u_{\perp}u_{\perp}}(-{\bm k},-\omega) \vert h_{u_{\perp}}({\bm k},\omega)\vert^2
\nonumber\\
&=& \frac{1}{\rho^2}\,\perpgradT \frac{1}{-i\omega + \DT k^2}\,\frac{-i\omega\,\nu k^2}{\omega^2 + \nu^2 k^4}\,\vert h_{u_{\perp}}({\bm k},\omega)\vert^2
\nonumber\\
&=& \frac{i}{\rho}\,\chi''_{Tu_{\perp}}({\bm k},\omega)\,\vert h_{u_{\perp}}({\bm k},\omega)\vert^2\ .
\label{eq:4.13}
\eea
Here we have used Eqs.~(\ref{eqs:4.3}) and (\ref{eqs:3.7}). The product $\delta{\tilde T}\,{\tilde u}_{\perp}$ yields the same
result, except for an overall minus sign. We see that the commutator correlation function $\chi''_{Tu_{\perp}}$ describes the bilinear response
of $\delta T\,u_{\perp}$ to the field $h_{u_{\perp}}$. Similarly, we have
\bea
\delta{\tilde T}({\bm k},\omega)\,\delta T (-{\bm k},-\omega) &=& X_{{\tilde T}u_{\perp}}({\bm k},\omega)\,X_{T u_{\perp}}(-{\bm k},-\omega) \vert h_{u_{\perp}}({\bm k},\omega)\vert^2
\nonumber\\
&=& \frac{1}{\rho^2} \perpgradT^2 \frac{1}{\omega^2 + \DT^2 k^4}\,\frac{i\omega \nu k^2}{\omega^2 + \nu^2 k^4}\,\vert h_{u_{\perp}}({\bm k},\omega)\vert^2
\nonumber\\
&=& \frac{i}{\rho}\,\chi_{TT}''^{\text{\,neq}}({\bm k},\omega)\,\vert h_{u_{\perp}}({\bm k},\omega)\vert^2\ ,
\label{eq:4.14}
\eea
with $\chi_{TT}''^{\text{\,neq}}$ the non-equilibrium part of the correlation function $\chi_{TT}''$ from Eq.~(\ref{eq:3.5b}). We finally observe
that the bilinear response of the product ${\tilde u}_{\perp}\,u_{\perp}$ is given by 
\bea
{\tilde u}_{\perp}({\bm k},\omega)\,u_{\perp}(-{\bm k},-\omega) &=& X_{{\tilde u}_{\perp} u_{\perp}}({\bm k},\omega)\,X_{u_{\perp} u_{\perp}}(-{\bm k},-\omega) \vert h_{u_{\perp}}({\bm k},\omega)\vert^2
\nonumber\\
&=& \frac{1}{\rho^2}\,\frac{i\omega\nu k^2}{\omega^2 + \nu^2 k^4}\,\vert h_{u_{\perp}}({\bm k},\omega)\vert^2\ ,
\nonumber\\
&=& \frac{i}{\rho}\,\chi_{u_{\perp}u_{\perp}}''({\bm k},\omega)\,\vert h_{u_{\perp}}({\bm k},\omega)\vert^2\ ,
\label{eq:4.15}
\eea
\end{widetext}
That is, although $\chi_{u_{\perp}u_{\perp}}''$ is, of course, equal to the spectrum of the {\em linear} response function
$X_{u_{\perp}u_{\perp}}$, it can also be written as a bilinear response. We also note that the products on the left-hand sides of
Eqs.~(\ref{eq:4.13}) - (\ref{eq:4.15}) all involve one observable ($\delta T$, or ${\tilde u}_{\perp}$) whose hydrodynamic equation 
contains a shear velocity relative to the external field, and one ($u_{\perp}$, or $\delta {\tilde T}$) whose equation contains an
absolute shear velocity.
 
The  Eqs.~(\ref{eq:4.13}) - (\ref{eq:4.15}) contain one of our main results: The commutator correlation functions in 
a NESS can be expressed as products of linear response functions, and hence are observable, as we anticipated
in the remarks after Eqs.~(\ref{eqs:3.7}). {Note that the Eqs.~(\ref{eq:4.13}) and (\ref{eq:4.14}) involve the
non-equilibrium parts of the commutator correlation functions only. In the limit of a vanishing temperature gradient
$\chi_{Tu_{\perp}}''$ vanishes and $\chi_{TT}''$ reduces to its equilibrium part that obeys the usual fluctuation-dissipation
theorem.}

The field $h_{u_{\perp}}$
can be realized by enforcing an initial shear flow on the system. Alternatively, one can eliminate the field in favor of initial conditions
and express the correlation functions in terms of the response of the system to initial perturbations $\delta T({\bm k},t=0)$
and $u_{\perp}({\bm k},t=0)$, see Eqs.~(\ref{eqs:B.8}) and Sec.~\ref{subsec:IV.C} below.

\subsection{Time evolution of external perturbations}
\label{subsec:IV.C}

Another way to probe the generalized rigidity of the system is via the relaxation of initial perturbations of the
temperature and the shear velocity. In this subsection we determine the relevant relaxation functions, and in
the following one we discuss the resulting propagation of a temperature perturbation. We note that the response
to initial conditions is equivalent to the response to external fields, and the relaxation functions can be expressed
in terms of the response functions, see Eqs.~(\ref{eqs:B.8}). 

\subsubsection{Hydrodynamic regime}
\label{subsubsec:IV.C.1}

We consider again the averaged simplified Navier-Stokes equations, but in the absence of external fields. We 
then have homogeneous equations
\bse
\label{eqs:4.16}
\bea
(\partial_t + \nu{\bm k}^2)\,u_{\perp}({\bm k},t) &=& 0\ ,
\label{eq:4.16a}\\
(\partial_t + \DT{\bm k}^2)\,\delta T({\bm k},t) + \perpgradT u_{\perp}({\bm k},t) &=& 0\ ,\qquad\quad
\label{eq:4.16b}
\eea
\ese
where we have transformed back to time space.
They are easily solved by means of a temporal Laplace transform defined as in Eq.~(\ref{eq:B.4a}). 
With $z$ as the complex frequency we find\cite{initial_conditions_footnote} 
\bse
\label{eqs:4.17}
\bea
u_{\perp}({\bm k},z) &=& M_{u_{\perp} u_{\perp}}({\bm k},z)\,u_{\perp}({\bm k},t=0)\ ,
\label{eq:4.17a}\\
\delta T({\bm k},z) &=& M_{Tu_{\perp}}({\bm k},z)\,u_{\perp}({\bm k},t=0) 
\nonumber\\
&&\hskip 30pt + M_{TT}({\bm k},z)\,\delta T({\bm k},t=0)\ .\qquad
\label{eq:4.17b}
\eea
\ese
The response or relaxation functions $M$ are related to the functions that describe the linear
response of the system to fields conjugate to the shear velocity and the temperature, respectively,
see Eqs.~(\ref{eqs:B.8}). For $\Im z > 0$, they are given by
\bse
\label{eqs:4.18}
\bea
M_{u_{\perp} u_{\perp}}({\bm k},z) &=& \frac{1}{-iz + \nu k^2}\ ,
\label{eq:4.18a}\\
M_{T u_{\perp}}({\bm k},z) &=& -\perpgradT\,\frac{1}{-iz + \nu k^2}\,\frac{1}{-i z + \DT k^2}\ ,
\nonumber\\
\label{eq:4.18b}\\
M_{TT}({\bm k},z) &=& \frac{1}{-i z + \DT k^2}\ .
\label{eq:4.18c}
\eea
\ese
Note that $M_{u_{\perp} u_{\perp}}$ and $M_{TT}$ are the same response functions as in equilibrium;\cite{local_equilibrium_footnote}
they are simple diffusion poles. $M_{T u_{\perp}}$ vanishes in equilibrium, but is nonzero in a NESS due to the coupling of the
shear velocity to the temperature in Eq.~(\ref{eq:4.16b}). It is a product of diffusion poles or, equivalently, a linear
combination of diffusion poles with a prefactor proportional to $1/k^2$. There is no response of the
shear velocity to an initial temperature perturbation since the temperature does not
couple into Eq.~(\ref{eq:4.16a}). Transforming back to time space we have
\bse
\label{eqs:4.19}
\bea
M_{u_{\perp} u_{\perp}}({\bm k},t) &=& e^{-\nu k^2 t}\ ,
\label{eq:4.19a}\\
M_{Tu_{\perp}}({\bm k},t) &=& \frac{\perpgradT}{(\nu - \DT) k^2}\left(e^{-\nu k^2 t} - e^{-\DT k^2 t}\right)\ ,
\nonumber\\
\label{eq:4.19b}\\
M_{TT}({\bm k},t) &=& e^{-\DT k^2 t}\ .
\label{eq:4.19c}
\eea
\ese

\subsubsection{Collisionless regime}
\label{subsubsec:IV.C.2}

In the collisionless regime we need to consider the kinetic equation (\ref{eq:2.47}) for the averaged
distribution function $\phi = \langle{\hat\phi}\rangle$. Here the procedure is more involved, and we
consider the equilibrium and non-equilibrium contributions to the response separately. 

\paragraph{Equilibrium contribution}
\label{par:IV.C.2a}

Consider Eq.~(\ref{eq:2.47}), averaged and in the absence of the temperature gradient. It can be
solved by a Fourier-Laplace transform:
\bse
\label{eqs:4.20}
\be
\phi({\bm p},{\bm k},z) = G_0({\bm p},{\bm k},z)\,\phi({\bm p},{\bm k},t=0)\ ,
\label{eq:4.20a}
\ee
with
\be
G_0({\bm p},{\bm k},z) = \frac{i}{z - {\bm k}\cdot{\bm v}_{\bm p}}\ .
\label{eq:4.20b}
\ee
\ese
Transforming back to time space yields
\bse
\label{eqs:4.21}
\be
\phi({\bm p},{\bm k},t) = G_0({\bm p},{\bm k},t)\,\phi({\bm p},{\bm k},t=0)\ ,
\label{eq:4.21a}
\ee
with
\be
G_0({\bm p},{\bm k},t) = e^{-i{\bm k}\cdot{\bm v}_{\bm p}t}
\label{eq:4.21b}
\ee
\ese
a real-time Green function. In terms of this solution the temperature fluctuations are given by\cite{Belitz_Kirkpatrick_2022, Kirkpatrick_Belitz_2022}
\bea
\delta T({\bm k},t) &=& \frac{1}{c_V}\,\frac{1}{V}\sum_{\bm p} w({\bm p})\,a_5({\bm p})\,\phi({\bm p},{\bm k},t)
\nonumber\\
&\equiv& \frac{1}{c_V}\,\left\langle a_5({\bm p})\vert \phi({\bm p},{\bm k},t)\right\rangle
\label{eq:4.22}
\eea
with $a_5({\bm p})$ from Eq.~(\ref{eq:A.1d}) and
\be
\langle g({\bm p})\vert h({\bm p})\rangle = \frac{1}{V}\sum_{\bm p} w({\bm p})\,g({\bm p})\,h({\bm p})
\label{eq:4.23}
\ee
the scalar product from Ref.~\onlinecite{Belitz_Kirkpatrick_2022} with $w$ the weight function defined in Eq.~(\ref{eq:2.42}).
If we multiply Eq.~(\ref{eq:4.21a}) from the left with $\langle a_5({\bm p})\vert$ and project the initial condition onto the temperature
by inserting a projector
\be
{\cal P}_5 = \frac{\vert a_5({\bm p})\rangle\langle a_5({\bm p})\vert}{\langle a_5({\bm p})\vert a_5({\bm p})\rangle}
\label{eq:4.24}
\ee
we obtain for the equilibrium part of the temperature fluctuations
\be
\delta T_{\text{eq}}({\bm k},t) = \frac{\left\langle a_5({\bm p})\big\vert e^{-i{\bm k}\cdot{\bm v}_{\bm p} t} a_5({\bm p})\right\rangle}{\langle a_5({\bm p})\vert a_5({\bm p})\rangle}\,\,\delta T({\bm k},t=0)\ .
\label{eq:4.25}
\ee

\paragraph{Non-equilibrium contribution}
\label{par:IV.C.2b}

For the non-equilibrium contribution we consider again the kinetic equation (\ref{eq:2.47}) without the fluctuating force,
but with the temperature-gradient term taken into account. The solution now reads
\begin{widetext}
\be
\phi({\bm p},{\bm k},t) = G_0({\bm p},{\bm k},t)\,\phi({\bm p},{\bm k},t=0)  - \frac{1}{T}\,a_s({\bm p}) \perpgradT \int_0^t d\tau\, G_0({\bm p},{\bm k},t-\tau)\,
      u_{\perp}({\bm k},\tau)\ .
\label{eq:4.26}
\ee
The transverse velocity $u_{\perp}$ is given by Eq.~(\ref{eq:A.4c}), which can be written
\be
u_{\perp}({\bm k},t) = \frac{1}{\rho}\,\langle a_{3}({\bm p})\vert\phi({\bm p},{\bm k},t)\rangle\ ,
\label{eq:4.27}
\ee
Multiplying Eq.~(\ref{eq:4.26}) from the left with $\langle a_5({\bm p})\vert$ we obtain the temperature fluctuation
in a NESS,
\be
\delta T({\bm k},t) = \frac{1}{c_V}\,\langle a_5({\bm p})\,e^{-i{\bm p}\cdot{\bm k}\,t} \vert \phi({\bm p},{\bm k},t=0)\rangle
 - \frac{ \perpgradT}{c_V T} \int_0^t d\tau\,\left\langle a_5({\bm p})\,e^{-i {\bm p}\cdot{\bm k}(t-\tau)}\vert a_s({\bm p})\right\rangle 
     u_{\perp}({\bm k},\tau)\ .
\label{eq:4.28}
\ee
The first term, if projected onto $\delta T({\bm k},t=0)$, is the equilibrium contribution given in Eq.~(\ref{eq:4.25}).
The second term is the non-equilibrium contribution. It is explicitly proportional to ${\bm\nabla}T$, so we can
take the equilibrium expression for $u_{\perp}$. The latter is obtained by multiplying Eq.~(\ref{eq:4.21a})
from the left with $\langle a_{3}({\bm p})\vert$. Projecting the initial condition on the transverse velocity,
as we did for the temperature in Eqs.~(\ref{eq:4.22}) - (\ref{eq:4.25}), we find
\be
u_{\perp}({\bm k},t) = \frac{1}{\rho}\,\left\langle a_3({\bm p})\big\vert e^{-i{\bm k}\cdot
   {\bm v}_{\bm p} t} a_3({\bm p})\right\rangle\, u_{\perp}({\bm k},t=0)\ .
\label{eq:4.29}
\ee
Here we have used the fact that the projection operations that lead to Eqs.~(\ref{eq:4.25}) and (\ref{eq:4.29}), respectively,
reflect the fact that we restrict the space of modes to the temperature and transverse velocity fluctuations.
Using Eq.~(\ref{eq:4.29}) in (\ref{eq:4.28}) yields the non-equilibrium contribution to $\delta T$:
\be
\delta T_{\text{neq}}({\bm k},t) =  \frac{\perpgradT}{\rho T}\,
 \frac{1}{V^2} \sum_{{\bm p}_1,{\bm p}_2} w({\bm p}_1)\,w({\bm p}_2)\,a_5({\bm p}_1) a_s({\bm p}_1) \left(a_3({\bm p}_2)\right)^2\,
  \,\frac{e^{-i{\bm p}_1 \cdot{\bm k} t} - e^{-i {\bm p}_2\cdot{\bm k} t}}{{\bm k}\cdot\left({\bm v}_{{\bm p}_1} - {\bm v}_{{\bm p}_2}\right)}
     \, u_{\perp}({\bm k},t=0)\ .
\label{eq:4.30}
\ee
Comparing Eqs.~(\ref{eq:4.30}) and (\ref{eq:4.25}) we see that the non-equilibrium contribution scales as the
equilibrium one with a $1/k$ prefactor, in analogy to the behavior in the hydrodynamic regime, where the
extra prefactor scaled as $1/k^2$.

Combining our results we now know the response functions defined by Eqs.~(\ref{eqs:4.17}) in the collisionless regime:
\bse
\label{eqs:4.31}
\bea
M_{u_{\perp} u_{\perp}}({\bm k},t) &=&  \frac{1}{\rho}\,\left\langle a_3({\bm p})\big\vert e^{-i{\bm k}\cdot
   {\bm v}_{\bm p} t} a_3({\bm p})\right\rangle\ ,
\label{eq:4.31a}\\
M_{T u_{\perp}}({\bm k},t) &=&  \frac{\perpgradT}{\rho T}\,
 \frac{1}{V^2} \sum_{{\bm p}_1,{\bm p}_2} w({\bm p}_1)\,w({\bm p}_2)\,a_5({\bm p}_1) a_s({\bm p}_1) \left(a_3({\bm p}_2)\right)^2\,
  \,\frac{e^{-i{\bm p}_1 \cdot{\bm k} t} - e^{-i {\bm p}_2\cdot{\bm k} t}}{{\bm k}\cdot\left({\bm v}_{{\bm p}_1} - {\bm v}_{{\bm p}_2}\right)}\ ,
\label{eq:4.31b}\\
M_{TT}({\bm k},t) &=&  \frac{1}{c_V T}\,\left\langle a_5({\bm p})\big\vert e^{-i{\bm k}\cdot{\bm v}_{\bm p} t} a_5({\bm p})\right\rangle\ .
\label{eq:4.31c}
\eea
\ese
\end{widetext}

\subsection{Propagation of a temperature perturbation}
\label{subsec:IV.D}

As another illustrative example of the dynamical consequences of the long-range correlations discussed in Sec.~\ref{sec:III}
we consider the response of the system to initial perturbations as expressed by Eqs.~(\ref{eqs:4.17}) - (\ref{eqs:4.19}).
For a classical fluid this problem has been discussed in Ref.~\onlinecite{Kirkpatrick_Belitz_Dorfman_2021}.
Suppose the temperature in a small sub-volume ${\cal V}$ is changed, at time $t=0$, by an amount $\delta T^{(0)}$,
and the transverse velocity $u_{\perp}$ is changed from zero to a value $u_{\perp}^{(0)}$. Let
${\cal V}$ be small enough that these initial perturbations can be described by spatial $\delta$-functions in a 
coarse-grained macroscopic description. The initial conditions in Fourier space are then independent of the 
wave number:\cite{anisotropy_footnote}
\bse
\label{eqs:4.32}
\bea
\delta T({\bm k},t=0) \equiv {\cal V}\,\delta T^{(0)}\ ,
\label{eq:4.32a}\\\,
u_{\perp}({\bm k},t=0) \equiv {\cal V}\,u_{\perp}^{(0)}\ ,
\label{eq:4.32b}
\eea
\ese
and we are interested in $\delta T({\bm x},t)$ at times $t>0$. As a measure of the propagation of the perturbation 
we consider the second spatial moments of $\delta T({\bm x},t)$, which we define by
\bea
\langle x_i^2\rangle &=& \frac{1}{T\,{\cal V}} \int d{\bm x}\,x_i^2\,\left[\delta T_{\text{eq}}({\bm x},t) + \vert\delta T_{\text{neq}}({\bm x},t)\vert\right]
\nonumber\\
&=& \langle x_i^2\rangle_{\text{eq}} + \langle x_i^2\rangle_{\text{neq}}
\label{eq:4.33}
\eea
where $x_1, x_2, x_3 \equiv x,y,z$. Here we have split $\langle x_i^2\rangle$ into an equilibrium contribution and a non-equilibrium
contribution, determined by Eqs.~(\ref{eq:4.19c}) and (\ref{eq:4.19b}), respectively. For the non-equilibrium contribution we take the absolute 
value, since $\delta T_{\text{neq}}$ can be either positive or negative, which has no physical significance.

\subsubsection{Hydrodynamic regime}
\label{subsubsec:IV.D.1}

In the hydrodynamic regime we use Eqs.~(\ref{eqs:4.19}). They are based on 
the Navier-Stokes equations, which have the same
form as in the classical case. We therefore obtain the same result as in Ref.~\onlinecite{Kirkpatrick_Belitz_Dorfman_2021}:
The propagation of the temperature perturbation is given by
\bea
\delta T({\bm k},t) &=& {\cal V}\, \delta T^{(0)}\,e^{-\DT k^2 t} 
\nonumber\\
&& + \frac{\perpgradT {\cal V}\,u_{\perp}^{(0)}}{(\nu - \DT) k^2} \left(e^{-\nu k^2 t} - e^{-\DT k^2 t}\right)\ .
\nonumber\\
\label{eq:4.34}
\eea
The first term is the equilibrium contribution, which has the usual diffusive form and is isotropic in ${\bm k}$-space. 
The second term is the non-equilibrium contribution, which is anisotropic. It is given by a linear combination of diffusive 
terms with a prefactor that scales as $1/k^2$. This is
consistent with the scaling of the non-equilibrium part of the commutator correlation $\chi_{TT}$, see
Eqs.~(\ref{eq:3.5b}) and (\ref{eq:3.6b}). Since the wave number squared scales as $k^2\sim 1/t$, this must lead
to an extra power of $t$, compared to the equilibrium contribution, in $\langle x_i^2\rangle$. Indeed, the calculation
yields, with ${\bm\nabla}T$ in the $z$-direction,
\bse
\label{eqs:4.35}
\bea
\langle x^2\rangle_{\text{eq}} &=& \langle y^2\rangle_{\text{eq}} = \langle z^2\rangle_{\text{eq}} = 2\DT \frac{\delta T^{(0)}}{T}\,t \ , \quad
\label{eq:4.35a}\\
\langle x^2\rangle_{\text{neq}} &=& \langle y^2\rangle_{\text{neq}} = \frac{1}{24}\,(\nu + \DT) \frac{t^2}{t_0}
\label{eq:4.35b}
\eea
where
\be
t_0 = T/\vert u_{\perp}^{(0)} \partial_z T\vert
\label{eq:4.35c}
\ee
is a time scale characteristic of the NESS, and
\be
\langle z^2\rangle_{\text{neq}} = 0\ .
\label{eq:4.35d}
\ee
\ese

The equilibrium contribution has the form expected for a perturbation that spreads diffusively and isotropically. 
The non-equilibrium contribution, in the plane perpendicular to ${\bm\nabla}T$, grows quadratically as a function of time, 
as expected from the above scaling argument. This anomalously fast propagation,
which is consistent with a propagating transport process rather than a diffusive one, 
is a consequence of the same generalized rigidity that causes the long-ranged correlations in the static
correlation functions, Eqs.~(\ref{eqs:3.6}). $\langle z^2\rangle_{\text{neq}}$ vanishes as a result of the
angular dependence of the non-equilibrium term in Eq.~(\ref{eq:4.34}).

\subsubsection{Collisionless regime}
\label{subsubsec:IV.D.2}

In the collisionless regime we must use Eqs.~(\ref{eqs:4.31}). The equilibrium contribution is again isotropic
and we find
\bea
\langle x^2\rangle_{\text{eq}} &=& \langle y^2\rangle_{\text{eq}} = \langle z^2\rangle_{\text{eq}} 
     = \frac{-1}{3T {\cal V}}\,{\bm\nabla}_{\bm k}^2\Big\vert_{{\bm k}=0} \delta T_{\text{eq}}({\bm k},t)
\nonumber\\
   &=& \frac{1}{3T}\,\delta T^{(0)}\,\frac{1}{\langle a_5({\bm p})\vert a_5({\bm p})\rangle}\,\langle a_5({\bm p})\vert {\bm v}_{\bm p}^2 a_5({\bm p})\rangle\, t^2\ .
   \nonumber\\
\label{eq:4.36}
\eea
In the low-temperature limit this becomes
\be
\langle r^2\rangle_{\text{eq}} \approx (\delta T^{(0)}/T)\,\vF^2\,t^2\ ,
\label{eq:4.37}
\ee
where $r^2 = {\bm x}^2$. This is the expected result for a ballistic mode with velocity $\vF$. 

The non-equilibrium contribution scales as the equilibrium one with a $1/k$ prefactor, see Eq.~(\ref{eq:4.30}).
Since $k \sim 1/t$ in the collisionless regime we again expect this to result
in an extra power of $t$ in the non-equilibrium contribution to $\langle r^2\rangle$. Indeed, the calculation yields
\bse
\label{eqs:4.38}
\bea
\langle x^2\rangle_{\text{neq}} &=& \langle y^2\rangle_{\text{neq}} = \frac{1}{3 c_V \rho T}\,
            \frac{1}{V^2} \sum_{{\bm p}_1,{\bm p}_2} w({\bm p}_1)\,w({\bm p}_2) \qquad
    \nonumber\\
 &&  \hskip -40pt \times\,a_5({\bm p}_1)\, a_s({\bm p}_1) \left(p_2^y\right)^2 
 \left(\frac{\left(v_{{\bm p}_1}^x\right)^3 - \left(v_{{\bm p}_2}^x\right)^3}{v_{{\bm p}_1}^x - v_{{\bm p}_2}^x}\right)\ \frac{t^3}{t_0}
\label{eq:4.38a}
\eea
with $t_0$ from Eq.~(\ref{eq:4.35c}), and
\be
\langle z^2\rangle_{\text{neq}} = 0\ .
\label{eq:4.38b}
\ee
\ese
We see that the temperature perturbation in a NESS propagates faster than ballistically as a result of the
generalized rigidity that is reflected in the long-ranged spatial correlations. $v_{\bm p}^x$ and $p^y$ in
Eq.~(\ref{eq:4.38a}) are the $x$ and $y$ components of ${\bm v}_{\bm p}$ and ${\bm p}$, respectively,
in an arbitrarily chosen cartesian coordinate systems. $\langle z^2\rangle_{\text{neq}}$ vanishes for the
same reason as in the hydrodynamic regime.

\subsection{Response versus fluctuations: The absolute size of the effect}
\label{subsec:IV.E}

A very interesting aspect of the response formulas derived in this section is that they allow for the
observation of the commutator correlation functions with a prefactor that is, (1) much larger than
the one in the corresponding fluctuation formulas, and (2) does not go to zero as $T\to 0$. To make
this point, consider the fluctuation formula (\ref{eq:3.8a}). $S^{\text{sym}}$ is directly observable by light
scattering, while $\chi$ is not, and the proportionality factor between the two is a microscopic energy,
viz., the temperature. We wish to compare this with the response formulas given by Eqs.~(\ref{eq:4.13})
- (\ref{eq:4.15}). Since the driving field $h_{u_{\perp}}$ is proportional to the shear velocity perturbation,
see Eqs.~(\ref{eqs:4.12}), a properly defined bilinear response function must be proportional to the anticommutator
correlation with the proportionality factor given by a macroscopic kinetic energy. To find the response function
analogous to the symmetrized correlation function, consider the linear response to an initial shear-velocity
perturbation, Eqs.~(\ref{eqs:4.17}). The initial condition $u_{\perp}({\bm k},t=0) = \int d{\bm x}\,e^{-i{\bm k}\cdot{\bm x}} u_{\perp}({\bm x},t=0)$
is a Fourier transform of a macroscopic velocity and hence scales as a macroscopic volume, and its
square scales as a volume squared. Now consider
the corresponding equations of motion (\ref{eqs:2.36}) for microscopic fluctuations and write them as
an initial-condition problem. The initial microscopic velocity $\hat{u}({\bm k},t=0)$ vanishes on average,
and the average of its symmetrized square is
\be
\frac{1}{2} \langle [{\hat u}_{\perp}({\bm k},t=0) , {\hat u}(-{\bm k},t=0)]_+\rangle = V\,T/\rho\ ,
\label{eq:4.39}
\ee
see Eq.~(\ref{eq:3.4b}), which scales as a volume. If we want to compare the response
formulas with the fluctuation formulas we therefore should divide the bilinear products in
Eqs.~(\ref{eq:4.13}) - (\ref{eq:4.15}) by a volume to compensate for this difference in the scaling
of the initial conditions with the volume. Accordingly, we define bilinear response functions analogous
to the correlation functions $S^{\text{sym}}$ as
\bse
\label{eqs:4.40}
\bea
\Sigma_{TT}({\bm k}) &=& \frac{-i}{V}\,\int \frac{d\omega}{\pi}\,\omega\,\delta{\tilde T}({\bm k},\omega)\,T(-{\bm k},-\omega)\ ,
\label{eq:4.40a}\\
\Sigma_{T u_{\perp}}({\bm k}) &=& \frac{-i}{V}\,\int \frac{d\omega}{\pi}\,\omega\,\delta T({\bm k},\omega)\,u_{\perp}(-{\bm k},-\omega),
\label{eq:4.40b}\\
\Sigma_{u_{\perp} u_{\perp}}({\bm k}) &=& \frac{-i}{V}\,\int \frac{d\omega}{\pi}\,\omega\,{\tilde u}_{\perp}({\bm k},\omega)\,u_{\perp}(-{\bm k},-\omega).\qquad\quad
\label{eq:4.40c}
\eea
\ese
The volume factor is motivated by the above considerations, and the frequency factor in the integrand
replaces one of the frequency integrations in the fluctuation formula (\ref{eq:3.8b}). From Eqs.~(\ref{eq:4.13}) - (\ref{eq:4.15})
together with (\ref{eqs:4.12}) we see that the energy that replaces $T$ in the fluctuation formula is
${\cal T} = \rho\,\frac{1}{V}\,\vert u_{\perp}({\bm k},t=0)\vert^2$. But the Fourier transform of the initial
shear velocity is on the order of $u_{\perp}({\bm k},t=0) \approx {\cal V}\,u_{\perp}^{(0)}$, with $\cal V$
the volume affected by the external perturbation and $u_{\perp}^{(0)}$ the magnitude of the externally
imposed shear velocity. We thus obtain response formulas
\bse
\label{eqs:4.41}
\bea
\Sigma_{TT}({\bm k}) &=& {\cal T}\,\chi_{TT}^{\text{neq}}({\bm k})
\label{eq:4.41a}\\
\Sigma_{T u_{\perp}}({\bm k}) &=& {\cal T}\,\chi_{T u_{\perp}}({\bm k})\ ,
\label{eq:4.41b}\\
\Sigma_{u_{\perp} u_{\perp}}({\bm k}) &=& {\cal T}\,\chi_{u_{\perp}u_{\perp}}({\bm k})\ .
\label{eq:4.41c}
\eea
\ese
Here $\chi_{TT}^{\text{neq}}({\bm k})$ is the non-equilibrium part of the static anticommutator correlation
function $\chi_{TT}$, Eq.~(\ref{eq:3.6b}), and $\chi_{T u_{\perp}}$ and $\chi_{u_{\perp} u_{\perp}}$ are
given by Eqs.~(\ref{eq:3.7f}) and (\ref{eq:3.4d}), respectively. The energy factor is
\be
{\cal T} = M\,(u_{\perp}^{(0)})^2 ({\cal V}/V)^2\ ,
\label{eq:4.42}
\ee
with $M$ the total mass of the fluid. $\cal T$ thus is the kinetic energy added to the fluid by the perturbation
times a factor of ${\cal V}/V$.

There are several remarkable aspects of this result. First, $\cal T$ is a macroscopic kinetic energy,
which is large compared to the microscopic energy $T$ (roughly the internal energy per particle) by
many orders of magnitude. Second, Eq.~(\ref{eq:4.41a}) provides a way to measure the non-equilibrium
part of $\chi_{TT}$ directly, with no equilibrium background. Third, Eq.~(\ref{eq:4.41b}) shows that the mixed
correlation function $\chi_{Tu_{\perp}}$ is observable. Finally, Eq.~(\ref{eq:4.41c}) provides a way to
measure $\chi_{u_{\perp} u_{\perp}}$ via a response experiment, even though it has the same form
as in equilibrium.

\section{Summary, and Discussion}
\label{sec:V}

In this section we first summarize our procedures and results, with an emphasis on how the various
sections of the paper are connected. We then discuss various points that received only cursory
mention in the main text.

\subsection{Summary}
\label{subsec:V.A}

The main purpose of this paper has been to consider the quantum analogs of the extraordinarily long-ranged correlations
that are known to generically exist in a classical fluid in a non-equilibrium steady state (NESS) characterized by a
constant temperature gradient, and to identify methods for observing them. The main challenges were, (1) the
different nature of soft modes that cause the long-range correlations in the hydrodynamic and collisionless
regimes, respectively, of a quantum fluid, (2) the necessity to distinguish between commutator and anticommutator
correlation functions, and (3) the lack of an established relation between correlation functions and response
functions. Conceptually, the most interesting and consequential point is the last one, which applies to classical
fluids as well as  quantum fluids.

In Sec.~\ref{sec:II} we considered a fluctuating quantum kinetic theory for fermions and used an adaptation of the classical
Chapman-Enskog method to derive fluctuating Navier-Stokes equations for the hydrodynamic regime. This was done mostly 
for completeness, as the Navier-Stokes equations must have the same structure as in a classical fluid. In order
to describe the collisionless regime we used the underlying kinetic equation. In equilibrium this theory
reduces to the one developed in Ref.~\onlinecite{Kirkpatrick_Belitz_2022}. We simplified both the
Navier-Stokes equations and the kinetic equation by keeping only the essential fluctuations, namely, the heat
mode and the shear velocity, and linearizing about the NESS.

In Sec.~\ref{sec:III} we calculated the temperature and shear velocity correlation functions in both the hydrodynamic
and collisionless regimes. In the hydrodynamic regime the temperature correlations display the same
long-rangedness as in a classical fluid. In the collisionless regime they are still long-ranged, but the singularity
is weaker due to the ballistic nature of the soft velocity modes, in agreement with an educated guess presented 
in the Introduction. The shear velocity fluctuation functions
have the same form as in equilibrium. All of these results rely on the premise that the correlations of the 
fluctuating forces are the same as in equilibrium. We will discuss this point in Sec.~\ref{subsec:V.B}.

In Sec.~\ref{sec:IV} we added a force conjugate to the shear velocity and calculated the related response
functions. We then showed that, for the problem under consideration, all of the commutator correlation
functions can be expressed in terms of products of response functions. That is, the long-range correlations
can be observed by probing the system's linear response to a macroscopic external perturbation, as an
alternative to measuring correlation functions. The resulting response formulas relate a product
of linear responses to a commutator correlation function via a macroscopic energy that is on the order of the
kinetic energy transferred to the fluid by the perturbation. By contrast, the energy in the corresponding
fluctuation formulas is the temperature, which is smaller by many orders of magnitude. We also
discussed another manifestation of the long-ranged
correlations, namely, the anomalous propagation of an initial temperature perturbation that is accompanied
by an initial shear perturbation. The temperature perturbation spreads faster than expected for diffusive
processes in the hydrodynamic regime, or ballistic ones in the collisionless regime, which is indicative of
the generalized rigidity that accompanies the long-ranged correlations.

\subsection{Discussion}
\label{subsec:V.B}

\subsubsection{The relative size of the effect}
\label{subsubsec:V.B.1}

\begin{table*}[t]
\label{tab:1}
\caption{Material Parameters Relevant for the Long-Range Correlations}
%\centering
%\vskip 6cm
\begin{ruledtabular}
\begin{tabular}{l|| l| l| l| l| l|| l| l}
%\multicolumn{8}{c}{Material Parameters Relevant for the Long-Range Correlations}                            \\
%\multicolumn{7}{c}{}              \\
%\hline
%\hline\\[-8pt]     
System           & Temperature  & Mass Density                   & Specific Heat               &  Thermal Diffusion                     & Kinematic                     & $k_{\text{mat}}$ (cm$^{-1})$    & $k^*$ (cm$^{-1})$           \\
                       & T (K)              & $\rho$ (g/cm$^3$)            &   $c_p/\rho$ (erg/g\,K)    &  Coefficient                                & Viscosity                       &  [1]                                               &  [2]                                    \\
                       &                      &                                           &                                      &  $\DT$ (cm$^2$/s)                    &  $\nu$ (cm$^2$/s)         &                                                  &                                         \\
\hline\hline\\[-8pt] 
n-hexane       & 298                 & 0.655\ [3]                         &  $2.264\times 10^7 $ [3]   &  $8.195\times 10^{-4}$ [3]    & $4.517\times 10^{-3}$ [3]   &  $3.93\times 10^7$              & $4,432$  \\                                                                                                                                                                                                        
\hline\\[-8pt]  
H$_2$O          & 298                 & 0.997  [4]                        &  $4.19\times 10^7 $  [4]   &  $1.45\times 10^{-3}$  [4]     & $0.89\times 10^{-2}$  [4]     &  $2.88\times 10^7$                & $3,795$  \\                                                                                                                                                                                                        
\hline\\[-10pt]   
\hline\\[-8pt]
Air                   & 300                &  $1.16\times 10^{-3}$ [4]  &  $1.00\times 10^7 $ [4]      &  $0.23 $    [4]                       & $0.160$   [4]                     &  $1.82\times 10^5$                      & $302$  \\                                                                                                                                                                                                        
\hline\\[-8pt]  
Ar                   & 300                 & $1.60\times 10^{-3}$ [4]   &  $5.21\times 10^6 $  [4]      &  $0.21$     [4]                      & $0.142$  [4]               &  $1.4\times 10^5$                  & $267$  \\                                                                                                                                                                                                        
\hline\\[-10pt]   
\hline\\[-8pt]
Liquid He3       & $1 $              &  $8.17\times 10^{-2}$ [5]  &  $1.43\times 10^7$  [6, 5]  & $\approx 7\times 10^{-4}$ [7, 6] &  $3.49\times 10^{-4}$ [8, 5]  & $4.4\times 10^6$             &  $1,485$       \\
\hline\\[-8pt]
Liquid Hg          & 298             &  $13.53$ [4]                   &  $1.40\times 10^6$ [4]        &   $4.3\times 10^{-2}$ [4]      &   $1.1\times 10^{-3}$  [4]  &  $4.69\times 10^5$                    &  $484$          \\
\hline\\[-8pt]
Liquid Ga          & 303             &  $6.09$   [9]                  &  $3.99\times 10^6$ [10]        &     $0.17$   [10]                      &  $3.23\times 10^{-3}$ [9]   &  $2.02\times 10^5$                 & $318$     \\
\hline\\[-10pt]
\hline\\[-8pt]   
Solid Al           & 298                &  $2.70$   [4]                 &  $8.97\times 10^6$ [4]           &   $0.98$   [4]                       &  $200$  [11]                         &  $3,900$                               & $60$          \\
\hline\\[-10pt]
\hline\\[-8pt]
Cold Atoms    & $\approx 5\times 10^{-8}$ [12] &  $1.7\times 10^{-9}$ [13]  &   $1.0\times 10^6$ [13]   &   $6.7\times 10^{-4}$ [13] &  $4.0\times 10^{-4}$ [13] &   270                   &   $12$             \\
\hline\\[-10pt]
\hline\\[-8pt]
Graphene      & 300                &  $2.25$ [14]                  &  $3\times 10^6$ [14]              &    $30$ [14]                          &   $500$ [15]                        &  $238$                                  &   $11$      \\
\hline\hline\\[-8pt]
References   &    \multicolumn{7}{l}{ [1] From Eq.~(\ref{eq:5.1c}); [2] For $k_{\text{exp}} = 0.5\,$cm$^{-1}$; [3] Ref.~\onlinecite{Li_et_al_1994}; [4] Ref.~\onlinecite{Rumble_2022} 
                                                          [5] Ref.~\onlinecite{Kerr_Taylor_1962}; [6] Ref.~\onlinecite{Roberts_Sydoriak_1955}.   }  \\
and                &    \multicolumn{7}{l}{ [7] From thermal conductivity data at various pressures, Ref.~\onlinecite{Anderson_et_al_1966}, extrapolated to $T=1$K.     }     \\
Notes            &    \multicolumn{7}{l}{ [8] Ref.~\onlinecite{Black_Hall_Thompson_1971}; [9] Ref.~\onlinecite{Spells_1936};  [10] At 313\,K, Ref.~\onlinecite{Schriempff_1973}.   }     \\
                     &   \multicolumn{7}{l}{  [11]  Calculated as $\nu = \vF\ell/5$ with $\vF \approx 2\times 10^8$cm/s and $\ell \approx 50$nm (Ref.~\onlinecite{Kojda_et_al_2015}). }     \\
                     &   \multicolumn{7}{l}{  [12] Ref.~\onlinecite{Stewart_Gaebler_Jin_2008};  [13] Calculated values, Eq.~(\ref{eq:5.3}), for noninteracting K$^{40}$ atoms with 
                                                                $\epsilonF \approx 0.5\mu$K.  } \\
                     &   \multicolumn{7}{l}{  [14] Measured valued from Ref.~\onlinecite{Pop_Vashney_Roy_2012}; [15] Calculated value from Ref.~\onlinecite{Principi_et_al_2016}.   }
\end{tabular}
\end{ruledtabular}
\vskip 1cm
\end{table*}

In Sec.~\ref{subsec:IV.E} we discussed the absolute size of the long-ranged correlations in the framework
of either spontaneous fluctuations, or the response to a macroscopic perturbation. Here we give a 
semi-quantitative discussion of the size of the effect relative to the equilibrium correlations
for various systems. We consider
the hydrodynamic regime and rewrite Eq.~(\ref{eq:3.6b}) as\cite{finite_size_footnote}
\bse
\label{eqs:5.1}
\be
\chi_{TT}({\bm k}) = \frac{T}{c_p}\,\left[1 + (k^*/k)^4\right]\ ,
\label{eq:5.1a}
\ee
where
\be
k^* = \left(k_{\text{exp}} k_{\text{mat}}\right)^{1/2}\ .
\label{eq:5.1a.1}
\ee
Here
\be
k_{\text{exp}} =  \perpgradT/T
\label{eq:5.1b}
\ee
is a wave number that can be controlled experimentally, and
\be
k_{\text{mat}} = \left( c_pT/\rho\DT(\nu+\DT)\right)^{1/2}
\label{eq:5.1c}
\ee
\ese
is a wave number that is material dependent. For $T_2 - T_1 \approx T$ (see Fig.~\ref{fig:1}) and $L \approx 1$cm,
one has $k_{\text{exp}} \approx 1\,$cm$^{-1}$. In the collisionless regime we have (see Eq.~(\ref{eq:3.13})
\be
\chi_{TT}({\bm k}) = \frac{1}{\NF}\,\frac{3}{\pi^2} \left[1 + s^{(a)}\,(T/\epsilonF)^2 (k_{\text{exp}}/k)^2\right]
\label{eq:5.2}
\ee
with $s^{(a)} \approx 13.8$ from Eq.~(\ref{eq:3.14b}).

In what follows we give rough estimates for the values of $k_{\text{mat}}$ and $k^*$ in various materials.
The relevant parameters are listed in Table~I}.

\paragraph{n-hexane:} This is the fluid used in the experiment in Ref.~\onlinecite{Li_et_al_1994}. $k_{\text{exp}}$ ranged
from $0.2$ to $0.5\,$cm$^{-1}$. With the parameters given in Table~I
one finds $k_{\text{mat}} \approx 4\times 10^7\,$cm$^{-1}$ and $k^* \approx 4,430\,$cm$^{-1}$. 
For the smallest wave number in that experiment,
$k = 1,607\,$cm$^{-1}$, this yields $(k^*/k)^4 \approx 58$. That is, the non-equilibrium 
contribution to $\chi_{TT}$ is larger than the equilibrium one by a factor of about $60$, in agreement with Fig.~6
in Ref.~\onlinecite{Li_et_al_1994} (see, e.g., Eq.~(8b) in Ref.~\onlinecite{Kirkpatrick_Belitz_Dorfman_2021} for
an expression of the factors $A_{\nu}$ and $A_T$ in terms of the material parameters).

The numbers for other classical liquids (e.g., water, see Table~I) at room temperature are similar.

\paragraph{Classical Gases:} The corresponding parameters for air and Ar, respectively, at room temperature yield
$k^* \approx 300$cm$^{-1}$. This smaller value compared to liquids is largely due to the transport coefficients being
larger by a factor of about $100$, which reflects the larger mean-free path.

\paragraph{Liquid He3:} In liquid He3 at $T=1\,$K the specific heat per mass and the transport coefficients are
comparable to those in classical liquids, which leads to $k^* \approx 1,500$. The smaller value of $k^*$ is due
to the lower temperature. 
For $k \approx k^*$ this still means that the non-equilibrium contribution is on the same order as the
equilibrium one. However, one needs to remember that a scattering experiment measures $S_{TT}$, which 
is suppressed by an overall factor of $T$ due to the low temperature, which makes a response experiment
attractive. In the collisionless regime the relative size of the non-equilibrium effect is still smaller, 
see Eq.~(\ref{eq:5.2}) ($\epsilonF \approx 1.5\,$K for He3.)

\paragraph{Solid Metals:} It is interesting to estimate $k^*$ in metals, even though $\chi_{TT}$ in metals cannot
be measured by light scattering. In a typical good metal the Fermi wave number is $\kF \approx 10^8\,$cm$^{-1}$, 
the Fermi velocity is $\vF \approx 10^8\,$cm/s, and $\epsilonF \approx 10^5$K. At temperatures low enough that 
electron-phonon scattering can be neglected, the electrons are well modeled as free fermions. The specific heat is
$c_p/\rho = \pi^2 T/2m\epsilonF$, with $m$ the fermion mass, and the diffusion coefficient and the shear velocity 
are $\DT= \vF^2\tau/3$ and $\nu = \vF^2\tau/5$, respectively, with $\tau = 2\epsilonF/\pi T^2$ the relaxation time. 
This yields
\be
k_{\text{mat}} = \pi^2\,\frac{3\sqrt{5}}{2^{11/2}}\,\left(\frac{T}{\epsilonF}\right)^3\,\kF \approx 0.15\,(T/\epsilonF)^3 \kF\ .
\label{eq:5.3}
\ee
With the parameters given above, and at, say, $T=1\,$K, $k_{\text{mat}}$ and $k^*$ are too small to be observable
by many orders of magnitude.

At higher temperatures (say, $T=300$K) the system can be considered a two-component fluid,
consisting of the electrons, which are still highly degenerate, and phonons. A typical electronic
mean-free path then is $\ell \approx 10^{-5}$cm.\cite{Kojda_et_al_2015} To estimate $k_{\text{mat}}$ we use the observed
values of the specific heat and the thermal diffusion coefficient; the values for Al are given in Table~I.
The kinematic viscosity one expects to be dominated by the electrons,
so we use $\nu \approx \vF\ell/5 \approx 200\,$cm$^2$/s. With $k_{\text{exp}} \approx 0.5\,$cm$^{-1}$ as in n-hexane this
yields $k^* \approx 60\,$cm$^{-1}$.

This relatively small value of $k^*$ brings up a complication that has to do with impurities.
The electron-impurity
scattering rate $1/\tau_{\text{i}} = \vF\ell_{\text{i}}$ gives the velocity a mass, and in order for the shear velocity to
remain diffusive one must have $1/\tau_{\text{e-i}} < \nu k^2 \approx \vF^2\tau_{\text{e-e}} k^2$, with
$\tau_{\text{e-e}}$ the electron-electron scattering time. This implies that the relation
\be
k^2 \ell_{\text{i}}\, \ell > 1
\label{eq:5.4}
\ee
is a necessary condition for the Navier-Stokes equations to be valid in an electron fluid. Together
with the general condition $k\ell < 1$ this means that the Navier-Stokes equations are applicable
only in a wave-number window
\be
1/\sqrt{\ell\,\ell_{\text{i}}} < k < 1/\ell\ .
\label{eq:5.5}
\ee
The existence of the window requires only $\ell_{\text{i}} > \ell$, but the lower bound is quite
restrictive even for ultraclean metals. A residual resistivity $\rho_{\text{i}} \approx 10^{-4}\mu\Omega\,$cm\cite{Ribot_et_al_1981}
corresponds to $\ell_{\text{i}} \approx 0.1\,$cm. With $\ell$ as above this yields the requirement
$k \agt 1,000\,$cm$^{-1}$. At lower temperatures the lower bound is smaller, but so is $k_{\text{mat}}$
and hence $k^*$. 

\paragraph{Liquid Metals:} The lower limit on the hydrodynamic window does not exist in a liquid
metal, which can be considered a two-component plasma, with the electrons again highly
degenerate. The parameters for mercury at room temperature and gallium just above the melting
point yield $k^* \approx 500\,$cm$^{-1}$ and $k^* \approx 300\,$cm$^{-1}$, respectively.

\paragraph{Cold Atoms:} In a typical fermionic cold-atom system, $\kF \approx 10^5\,$cm$^{-1}$
and $T/\epsilonF \approx 0.1$.\cite{Stewart_Gaebler_Jin_2008} Equation~(\ref{eq:5.3}) then yields 
$k_{\text{mat}} \approx 250\,$cm$^{-1}$ and $k^* \approx 10\,$cm$^{-1}$, see Table~I. The low 
density of these systems leads to a small $\kF$, which suppresses the effect.     

\paragraph{2-d Electron Systems:} An interesting 2-d electron system is provided by graphene, where the
hydrodynamic condition (\ref{eq:5.4}) is easier to satisfy than in metals. Using the observed values for the
specific heat of graphene at $T\approx 300\,$K \cite{Pop_Vashney_Roy_2012} and the calculated value
for the kinematic viscosity\cite{Principi_et_al_2016} one finds
$k_{\text{mat}} \approx 240\,$cm$^{-1}$, and $k^* \approx 11\,$cm$^{-1}$, which is about the same as
in a cold-atom system. 

\bigskip
These rough estimates indicate that the effect in quantum systems becomes sizable only at substantially
smaller wave numbers than in classical ones, with liquid He3 and liquid metals the most promising
systems. This makes the manifestations of the effect as a response
to {\em macroscopic} perturbation that were discussed in Secs.~\ref{subsec:IV.B} - \ref{subsec:IV.E} 
attractive for experimental purposes. In this context we mention again that the response formula
(\ref{eq:4.41a}) allows for measurements of the non-equilibrium part of the correlation function $\chi_{TT}$
separately, with no background provided by the equilibrium contributions.

\subsubsection{General discussion}
\label{subsubsec:V.B.2}

We finally discuss in more detail several aspects of our procedure and our results.

\begin{enumerate} [wide, labelindent=10pt, label=\arabic*)]
\item
An important question that has not been emphasized in the past is whether the long-ranged
correlations we have discussed throughout this paper are due to thermal fluctuation effects, or
whether they are more generic and reflect a generalized rigidity, in the sense of Ref.~\onlinecite{Anderson_1984},
that is inherent to the NESS. Recent work on classical fluids\cite{Kirkpatrick_Belitz_Dorfman_2021} suggested 
the latter. This conclusion is supported by the fact that the correlation functions are long-ranged even though 
the correlations of the fluctuating forces are not, see Eqs.~(\ref{eqs:2.38}).\cite{LTT_footnote}
This is consistent with the fact that the results for classical fluids are the same irrespective of whether they are 
obtained, as in the original derivations,\cite{Kirkpatrick_Cohen_Dorfman_1982a, Kirkpatrick_Cohen_Dorfman_1982c} 
by using kinetic theory and mode-coupling theory, or by using fluctuating hydrodynamics with short-ranged
random-force correlations.\cite{Ronis_Procaccia_1982, fluctuating_forces_footnote} This implies that the long-rangedness is encoded
in the averaged hydrodynamic equations. Indeed, as we have shown in Sec.~\ref{sec:IV} of the present
paper, the solution of the initial-value problem for the latter contains the long-rangedness and the
related generalized rigidity. 

\item
A related issue is the use of random-force correlations that are the same as in local equilibrium. 
Strictly speaking this represents an assumption, although various plausibility arguments have
been given, see Refs.~\onlinecite{Ronis_Procaccia_Machta_1980, Kirkpatrick_Dorfman_2015}.
The calculation of the correlation functions nevertheless reveals strong long-ranged correlations
and associated generalized rigidity. This is true both classically and quantum mechanically, and
it indicates that long-ranged correlations are an inherent aspect of hydrodynamics in a NESS, see
above. The experimental results in classical fluids are in very good agreement with the theory,
which lends further credence to the assumption. Measuring the response functions, in addition to the correlation functions,
would provide another check on the validity of the assumption. The development
in Sec.~\ref{sec:IV} makes such a check possible; previously a direct connection between 
long-ranged correlation functions and some type of response theory had been lacking.

\item
A major obstacle for establishing a relation between correlations and response had been
the absence, in any system that is not in thermodynamic equilibrium,
of a simple fluctuation-dissipation theorem that relates correlation functions to response functions. 
Specifically, commutator correlation functions no longer are equal to 
response functions and their physical meaning is {\em a priori} unclear. Substantial work has
been done on fluctuations in systems far from equilibrium (see, Refs.~\onlinecite{Sevick_et_al_2008,
Gaspard_2022} and references therein), on non-equilibrium linear response,\cite{Baiesi_Maes_2013}
and on the relation between these topic.\cite{Gaspard_2022, Andrieux_Gaspard_2004, Maes_2020}
However, there has been no prescription for probing fluctuations via the system's response to 
external perturbations. We have studied a very 
simple non-equilibrium state, viz., a fluid in a NESS characterized by a constant temperature gradient,
and have employed various simplifications.\cite{approximations_footnote}
For this system we have shown, in Sec.~\ref{sec:IV}, that there still is a connection between the 
correlation functions and the response functions: The former are related to the bilinear response of 
products of observables to a field conjugate to the shear velocity. The fluid's response to
external perturbations thus contains the same information about the generalized rigidity as
the correlation functions. This unexpected relation between 
correlations and response opens an alternative way for experimentally probing the long-ranged 
correlations. It is important to keep in mind that the relation holds due to the structure of
the hydrodynamic equations in the NESS, and in particular to the fact that the shear-velocity
correlation functions still describe the linear response to a conjugate field, as they do in equilibrium, see the remarks
after Eqs.~(\ref{eqs:3.7}). We also note that it is true for causal functions that represent
simple resonances, in particular diffusion poles, but not for causal functions in general. 

\item
It is illustrative to compare Eqs.~(\ref{eqs:4.2}, \ref{eqs:4.3}) and (\ref{eqs:3.3}). The expression
for $\delta T$ in Eq.~(\ref{eq:4.2b}), with the response function given by Eq.~(\ref{eq:4.3b}), 
is the same as the non-equilibrium contribution in the first
line of (\ref{eq:3.3b}), with the fluctuating force $P_{\perp}$ replaced by a macroscopic driving
force proportional to the field $h_{u_{\perp}}$. This makes it plausible that the non-equilibrium
part of the commutator correlation function $\chi_{TT}''$ is related to the bilinear response of
a product of two $\delta T$ factors, as Eq.~(\ref{eq:4.14}) demonstrates. In some sense this
is a more plausible structure than in equilibrium, where the commutator correlation function
is a {\em linear} response function.

\item
The light-scattering experiments that have been used to experimentally confirm the long-range correlations 
in classical fluids in a NESS, see Ref.~\onlinecite{Sengers_Ortiz_Kirkpatrick_2016} and references
therein, are difficult since they require very small scattering angles. This is even more
relevant in quantum fluids, since the fluctuation effects become weaker with decreasing temperature. 
Our predictions in Sec.~\ref{subsec:IV.B} open another route to measuring the long-ranged correlations,
namely, via the response of the system to macroscopic perturbations. Alternatively, they can be probed
via the propagation of an initial temperature pulse that is accompanied by an initial perturbation of the
shear velocity, as was discussed for classical fluids in Ref.~\onlinecite{Kirkpatrick_Belitz_Dorfman_2021}
and in the present context in Secs.~\ref{subsec:IV.C}, \ref{subsec:IV.D}. Such experiments
will also test the prediction of the theory that, while the fluctuation effects become weaker with
decreasing temperature, the generalized rigidity does not. 

\item
Light-scattering experiments actually measure the correlation function of the fluctuations of the
dielectric function, which are largely proportional to the density fluctuations (the contribution of
the temperature fluctuations is small), see, e.g., Appendix A.4 of Ref.~\onlinecite{Forster_1975}. 
The fluctuations of interest, namely, those of the temperature or
the entropy per particle, are reflected in the central Rayleigh peak of the density-density correlation 
function. The sound modes are reflected in the Brillouin peaks. Neglecting the latter, as we
have done, changes the specific heat $c_V$ in Eqs.~(\ref{eqs:1.1}) and (\ref{eq:1.2}) to $c_p$, 
as the sound modes give an  additional contribution to the sum rule; see Ref.~\onlinecite{Forster_1975}
for a discussion. This is why the equilibrium contributions to Eqs.~(\ref{eqs:3.5}) contain $c_p$.

\item
Our results hold {\em a fortiori} for charged Fermi liquids, i.e., for conduction 
electrons in metals, since the Coulomb interaction renders massive the longitudinal sound modes,
which we have neglected, but has no effect on the other soft modes. In particular, for a metal in the
collisionless regime our results hold without the caveat that the pressure fluctuations, which we
have neglected, have an effect that is qualitatively the same as the effect of the shear modes, which we have
kept, see the remarks after Eqs.~(\ref{eqs:3.15}). However, all other zero modes, to the extent that they exist 
(this depends on the values of the Landau Fermi-liquid parameters), still yield contributions that scale the 
same as those from transverse zero sound.

\end{enumerate}

\appendix

\section{Fluctuations of observables, and hydrodynamic modes}
\label{app:A}

Here we recall how various observables are represented in the kinetic theory. For derivations see Ref.~\onlinecite{Belitz_Kirkpatrick_2022}. The notation
is the same as in that reference unless noted otherwise.

Define the functions
\bse
\label{eqs:A.1}
\bea
a_1({\bm p}) &=& 1\ ,
\label{eq:A.1a}\\
a_2({\bm p}) &=& \hat{\bm k}\cdot{\bm p}\ ,
\label{eq:A.1b}\\
a_{3,4}({\bm p}) &=& \hat{\bm k}_{\perp}^{(1,2)}\cdot{\bm p}\ ,
\label{eq:A.1c}\\
a_5({\bm p}) &=& \epsilon_p - \mu + T\left(\frac{\partial\mu}{\partial T}\right)_{N,V}\ ,
\label{eq:A.1d}
\eea
as in Ref.~\onlinecite{Belitz_Kirkpatrick_2022}, and in addition
\bea
a_p({\bm p}) &=& \frac{1}{c_V}\left(\frac{\partial p}{\partial T}\right)_{N,V} a_5({\bm p}) + \left(\frac{\partial p}{\partial n}\right)_{T,V} a_1({\bm p})\ ,\hskip 25pt
\label{eq:A.1e}\\
a_s({\bm p}) &=&  \epsilon_p - \mu - sT/n = a_5({\bm p}) - \frac{T}{n}\left(\frac{\partial p}{\partial T}\right)_{N,V} a_1({\bm p})\ .
\nonumber\\
\label{eq:A.1f}
\eea
\ese
Here $\hat{\bm k}$, $\hat{\bm k}_{\perp}^{(1)}$, and $\hat{\bm k}_{\perp}^{(2)}$ form a right-handed orthogonal system of unit vectors,
with $\hat{\bm k} = {\bm k}/k$ the unit wave vector. All other quantities are as defined in Sec.~\ref{sec:II}.
\begin{figure}[b]
\includegraphics[width=6cm]{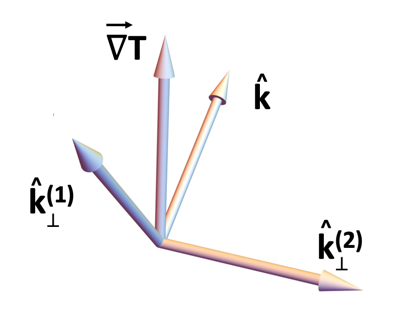}
\caption{The three vectors spanning ${\bm k}$-space, and the temperature gradient.}
\label{fig:3}
\end{figure}
For our purposes it is convenient to choose a coordinate system such that $\hat{\bm k}_{\perp}^{(2)}$ is orthogonal to both
$\hat{\bm k}$ and $\bm\nabla T$, and $\hat{\bm k}_{\perp}^{(1)}$ lies in the plane spanned by $\hat{\bm k}$ and $\bm\nabla T$,
see Fig.~\ref{fig:3}. With this convention only ${\hat{\bm k}}_{\perp}^{(1)} \equiv {\hat{\bm k}}_{\perp}$ and 
$u_{\perp} = \hat{\bm k}_{\perp}\cdot{\bm u}$ contribute to the scalar product ${\bm\nabla}T\cdot{\bm u}_{\perp}$. 
If we choose ${\bm\nabla} T$ to point in the $z$-direction we have explicitly
\bse
\label{eqs:A.2}
\bea
{\hat{\bm k}}_{\perp}^{(1)} &=& \frac{1}{k\sqrt{k_x^2 + k_y^2}}\,(-k_x k_z,-k_y k_z,k_x^2 + k_y^2)\ ,\qquad
\label{eq:A.2a}\\
{\hat{\bm k}}_{\perp}^{(2)} &=& \frac{1}{\sqrt{k_x^2 + k_y^2}}\,(k_y,-k_x,0)\ .
\label{eq:A.2b}
\eea
\ese

We will also need the normalizations\cite{Belitz_Kirkpatrick_2022}
\bse
\label{eqs:A.3}
\bea
\langle a_1({\bm p})\vert a_1({\bm p})\rangle &=& (\partial n/\partial\mu)_{T,V}\ ,
\label{eq:A.3a}\\
\langle a_2({\bm p})\vert a_2({\bm p})\rangle &=& \langle a_3({\bm p})\vert a_3({\bm p})\rangle = \langle a_4({\bm p})\vert a_4({\bm p})\rangle = \rho\ ,
\nonumber\\
\label{eq:A.3b}\\
\langle a_5({\bm p})\vert a_5({\bm p})\rangle &=&c_V T\ ,
\label{eq:A.3c}\\
\langle a_s({\bm p})\vert a_s({\bm p})\rangle &=&c_p T\ ,
\label{eq:A.3d}
\eea
\ese
with $\langle\ \vert\ \rangle$ the scalar product from Ref.~\onlinecite{Belitz_Kirkpatrick_2022} (see also
Eq.~(\ref{eq:4.23}). Equations~(\ref{eq:A.3a}) and (\ref{eq:A.3b}) hold for noninteracting electrons only.

The fluctuations of the particle number density $n$, the longitudinal fluid velocity $u_{\text{L}}$, the relevant
component of the transverse velocity, $u_{\perp}$, and the temperature $T$ are given by
\bse
\label{eqs:A.4}
\bea
\delta n({\bm x},t) &=& \frac{1}{V}\sum_{\bm p} a_1({\bm p})\,\delta f_{\bm p}({\bm x},t)\ ,
\label{eq:A.4a}\\
\delta u_{\text{L}}({\bm x},t) &=& \frac{1}{\rho}\,\frac{1}{V}\sum_{\bm p} a_2({\bm p})\,\delta f_{\bm p}({\bm x},t)\ ,
\label{eq:A.4b}\\
\delta u_{\perp}({\bm x},t) &=& \frac{1}{\rho}\,\frac{1}{V}\sum_{\bm p} a_3({\bm p})\,\delta f_{\bm p}({\bm x},t)\ ,
\label{eq:A.4c}\\
\delta T({\bm x},t) &=& \frac{1}{c_V}\,\frac{1}{V}\sum_{\bm p} a_5({\bm p})\,\delta f_{\bm p}({\bm x},t)\ ,
\label{eq:A.4d}
\eea
and those of the pressure $p$ and the entropy per particle $s/n$ by
\bea
\delta p({\bm x},t) &=& \frac{1}{V}\sum_{\bm p} a_p({\bm p})\,\delta f_{\bm p}({\bm x},t)\ ,
\label{eq:A.4e}\\
\delta (s/n)({\bm x},t) &=& \frac{1}{Tn}\, \frac{1}{V}\sum_{\bm p} a_s({\bm p})\,\delta f_{\bm p}({\bm x},t)\ .
\label{eq:A.4f}
\eea
The complete transverse velocity fluctuation is
\be
\delta {\bm u}_{\perp}({\bm x},t) = \frac{1}{\rho}\,\frac{1}{V}\sum_{\bm p} \left[ a_3({\bm p}){\hat{\bm k}}_{\perp}^{(1)} + a_4({\bm p}){\hat{\bm k}}_{\perp}^{(2)}\right]  \delta f_{\bm p}({\bm x},t)\ .
\label{eq:A.4g}
\ee
\ese

Of these fluctuations, only $\delta (s/n)$ and $\delta {\bm u}_{\perp}$ are hydrodynamic modes, viz., the heat mode
and the shear modes, respectively. They are all diffusive. The remaining hydrodynamic
modes are two propagating longitudinal sound modes, given by the linear combinations
\be
\delta p({\bm x},t) \pm c_1 \rho\,\delta u_{\text{L}}({\bm x},t)
\label{eq:A.5}
\ee
with $c_1$ the speed of (first) sound. The temperature fluctuations can be written as linear combinations of entropy
fluctuations $\delta(s/n)$ and pressure fluctuations $\delta p$ by combining Eqs.~(\ref{eq:A.1d}), (\ref{eq:A.1e}), and
(\ref{eq:A.1f}). After using some thermodynamic identities we find\cite{coefficients_footnote}
\bse
\label{eqs:A.6}
\be
a_5({\bm p}) = \frac{c_V}{c_p}\,\left[ a_s({\bm p}) - \frac{T}{n}\left(\frac{\partial n}{\partial T}\right)_{p,V} a_p({\bm p})\right]
\label{eq:A.6a}
\ee
or
\be
\delta T({\bm x},t) = \frac{Tn}{c_p}\,\delta(s/n)({\bm x},t) - \frac{T}{n c_p}\left(\frac{\partial n}{\partial T}\right)_{p,V} \delta p({\bm x},t)\ .
\label{eq:A.6b}
\ee
\ese
Note that the two contributions to the longitudinal sound modes in Eq.~(\ref{eq:A.5}) are mutually orthogonal, so Eq.~(\ref{eq:A.6b})
expresses the temperature fluctuation as a linear combination of two hydrodynamic modes plus a contribution that is orthogonal
to both of these modes. Similarly, we can write density fluctuations as linear combinations of entropy and pressure fluctuations. 
We find\cite{coefficients_footnote}
\bea
\delta n({\bm x},t) &=& \frac{Tn}{c_p} \left(\frac{\partial n}{\partial T}\right)_{p,V} \delta(s/n)({\bm x},t) 
\nonumber\\
&& + \frac{c_V}{c_p} \left(\frac{\partial n}{\partial p}\right)_{T,V} \delta p({\bm x},t)\ .
\label{eq:A.7}
\eea

\section{Correlation functions, response functions, the fluctuation-dissipation theorem, {and the initial-value problem}}
\label{app:B}

Here we recall the definitions of various correlation and response functions, with emphasis on a crucial difference
between equilibrium and non-equilibrium systems. See Refs.~\onlinecite{Kadanoff_Martin_1963} and \onlinecite{Forster_1975}
for detailed discussions of the equilibrium case. In contrast to the main text, we explicitly keep $\hbar$. {We also
discuss the equivalence between linear response to an external field and an initial-value problem.}

\subsection{{Correlation functions}}
\label{app:B.1}

Let ${\hat A}_i({\bm x},t)$ ($i=1,2,\ldots$) be observables. Then the van Hove function
\be
S_{A_iA_j}({\bm x},{\bm x}';t-t') = \langle\delta{\hat A}_i({\bm x},t)\,\delta{\hat A}_j({\bm x}',t')\rangle
\label{eq:B.1}
\ee
with $\delta{\hat A}_i = {\hat A}_i - \langle{\hat A}_i\rangle$ 
describes correlations between the fluctuations of two observables at different points in space-time. 
Related correlation functions are the symmetrized or anticommutator correlation function
\bse
\label{eqs:B.2}
\be
\Ssym_{A_iA_j}({\bm x},{\bm x}';t-t') = \frac{1}{2}\langle[\delta{\hat A}_i({\bm x},t),\delta{\hat A}_j({\bm x}',t')]_+\rangle
\label{eq:B.2a}
\ee
and the antisymmetrized or commutator correlation function
\be
\chi_{A_iA_j}''({\bm x},{\bm x}';t-t') = \frac{1}{2\hbar}\langle[{\hat A}_i({\bm x},t),{\hat A}_j({\bm x}',t')]_-\rangle\ .
\label{eq:B.2b}
\ee
Here $[\ ,\ ]_{\pm}$ denotes an anticommutator and commutator, respectively, and $\langle \ldots\rangle$ 
indicates a quantum mechanical expectation value plus a statistical mechanics average. 
The relation between the  symmetrized and antisymmetrized correlation functions 
in non-equilibrium systems is not known in general. However, 
for the particular NESS we consider in this paper one can, to leading order in the effects of the
temperature gradient, replace the temperature by its spatially averaged value everywhere except 
in the crucial coupling term between the temperature gradient and the shear velocity. Within this
approximation, the temporal Fourier transforms two correlation functions are related by the same factor as in equilibrium,
\be
\Ssym_{A_iA_j}({\bm x},{\bm x}';\omega) = \hbar\,\coth(\hbar\omega/2T)\,\chi_{A_iA_j}''({\bm x},{\bm x}';\omega)\ .
\label{eq:B.2c}
\ee
\ese
We stress, however, that this does {\em not} imply that $\chi''$ is a linear response function.

\subsection{{Response functions, and the fluctuation-dissipation theorem}}
\label{app:B.2}

Let $h_{A_i}$ be an external field conjugate to ${\hat A}_i$. Then the response function
$X_{A_iA_j}$ is defined via
\be
\delta\langle {\hat A}_i({\bm x},t)\rangle = \int_{-\infty}^t dt'\,X_{A_iA_j}({\bm x},{\bm x'},t-t')\,h_{A_j}({\bm x}',t')\ .
\label{eq:B.3}
\ee
Let 
\bse
\label{eqs:B.4}
\bea
{X_{A_iA_j}({\bm x},{\bm x}';z)} &=& {\pm\int_{-\infty}^{\infty} dt\,\Theta(\pm t)\,e^{izt}\,X_{A_iA_j}({\bm x},{\bm x'},t)}
\nonumber\\
&& { (\pm\ \text{for}\ \text{Im}(z)  \genfrac{}{}{0pt}{2}{>}{<} 0)\ ,}
\label{eq:B.4a}
\eea
{with $\Theta$ the step function,}
be the temporal Laplace transform of $X_{A_iA_j}$, with $z$ the complex frequency,{\cite{Laplace_trafo_footnote}} and
\bea
X_{A_iA_j}''({\bm x},{\bm x}';\omega) &=& \frac{1}{2i}\,[X_{A_iA_j}({\bm x},{\bm x}';\omega+i0) 
\nonumber\\
&& \hskip 20pt - X_{A_iA_j}({\bm x},{\bm x}';\omega-i0)]\qquad
\label{eq:B.4b}
\eea
its {spectral density, which is related to the Laplace transform via the Hilbert-Stieltjes relation}
\be
{X_{A_iA_j}({\bm x},{\bm x}';z) = \int_{-\infty}^{\infty} \frac{d\omega}{\pi}\,\frac{X_{A_iA_j}''({\bm x},{\bm x}';\omega)}{\omega - z}\ .}
\label{eq:B.4c}
\ee
{The response functions in Eqs.~(\ref{eqs:4.2}) and (\ref{eqs:4.3}) are}
\be
{X_{A_iA_j}({\bm x},{\bm x}';\omega) = \lim_{\epsilon\to 0+} X_{A_iA_j}({\bm x},{\bm x}'; z = \omega + i\epsilon)\ .}
\label{eq:B.4d}
\ee
\ese
{The spectral densities $X_{A_iA_j}''$ determine, at bilinear order in the external fields, the energy dissipated due to the work
done by the fields.\cite{Forster_1975} In equilibrium, $X_{A_iA_j}'' = \chi_{A_iA_j}''$, and hence Eq.~(\ref{eq:B.2c}) implies}
\be
\Ssym_{A_iA_j}({\bm x},{\bm x}';\omega) = \hbar\,\coth(\hbar\omega/2T)\,X_{A_iA_j}''({\bm x},{\bm x}';\omega)\ .
\label{eq:B.5}
\ee
This is a manifestation of the fluctuation-dissipation theorem,\cite{Nyquist_1928, Callen_Welton_1951}
which relates correlations of the equilibrium fluctuations, described by $\Ssym_{A_iA_j}$, to the energy dissipated,
described by $X_{A_iA_j}''$. It holds only in equilibrium, {where Eq.~(\ref{eq:B.2c}) is exact, and  $X_{A_iA_j}'' = \chi_{A_iA_j}''$.
In a non-equilibrium system $\chi_{A_iA_j}''$ and $X_{A_iA_j}''$ are in general not the same, and there is no simple
general relation between correlation functions and response functions.}

\subsection{{Linear response as an initial-value problem}}
\label{app:B.3}

Now consider a spatially homogeneous system for simplicity, perform a spatial Fourier transform {in
Eq.~(\ref{eq:B.3}), and consider an external field that is adiabatically switched on at time $t=-\infty$
and discontinuously switched off at $t=0$:}
\be
{h_{A_j}({\bm k},t) = h_{A_j}({\bm k})\,e^{\epsilon t}\,\Theta(-t)\ ,}
\label{eq:B.6}
\ee
{where $\epsilon>0$ is infinitesimal and positive. Then at time $t=0$ the field leads to a nonzero 
expectation value of ${\hat A}_i$ given by}
\be
{\delta\langle {\hat A}_i({\bm x},t=0)\rangle = X_{A_iA_j}({\bm k})\,h_{A_j}({\bm k})\ ,}
\label{eq:B.7}
\ee
{with $X_{A_iA_j}({\bm k}) = X_{A_iA_j}({\bm k},z=0)$ the static response functions. Now one can
use Eq.~(\ref{eq:B.7}) to eliminate the fields from Eq.~(\ref{eq:B.3}).} Let $X({\bm k},z)$ be the matrix of
response functions $X_{A_iA_j}({\bm k},z)$, and let $X({\bm k}) = X({\bm k},z=0)$ be the corresponding 
matrix of static response functions. Then a straightforward calculation yields
\bse
\label{eqs:B.8}
\be
\delta\langle {\hat A}_i({\bm k},z)\rangle = \sum_j M_{ij}({\bm k},z)\, \delta\langle {\hat A}_j({\bm k},t=0)\rangle
\label{eq:B.8a}
\ee
where
\be
M_{ij}({\bm k},z) = \frac{1}{iz}  \left[X({\bm k},z) X^{-1}({\bm k}) - \mathbb{1}\right]_{ij}\,
\label{eq:B.8b}
\ee
\ese
with $\mathbb{1}$ the unit matrix. {This result, and its derivation, is verbatim the same
as for equilibrium systems, where $X = \chi$, see, e.g., Sec.~3.1 in Ref.~\onlinecite{Forster_1975}.
It holds in complete generality as long as $X$ is taken to be the response function; with $X$ replaced
by $\chi$ it holds only in equilibrium.}

\subsection{{Non-hydrodynamic initial condition for the shear velocity}}
\label{app:B.4}

{In the case of the initial condition for the shear velocity, Eq.~(\ref{eq:4.17a}), the following
complication occurs, which was noted in Ref.~\onlinecite{Kirkpatrick_Belitz_2023a}. The pressure
tensor for the fluid is}
\bse
\label{eqs:B.9}
\be
{P_{ij}({\bm k},t) = \delta_{ij}\,p({\bm k},t) - \sigma_{ij}({\bm k},t)\ ,}
\label{eq:B.9a}
\ee
{with $p({\bm k},t)$ the hydrostatic pressure and}
\bea
{\sigma_{ij}({\bm k},t)} &=& {-\eta\,\Bigl[-i k_i u_j({\bm k},t) - i k_j u_i({\bm k},t)}
\nonumber\\
&&\hskip 50pt {+ \frac{2}{3}\,\delta_{ij}\,i {\bm k}\cdot{\bm u}({\bm k},t)\Bigr]}
\label{eq:B.9b}
\eea
\ese
{the stress tensor. Note that Eq.~(\ref{eq:B.9b}) is just a fancy version of Hooke's law: the generalized force (stress) equals
a generalized elastic constant (shear viscosity) times the generalized displacement (strain). Consequently, an initial shear
velocity $u_{\perp}({\bm k},t=0)$ leads to a nonzero initial shear stress, viz.}
\be
{\sigma_{\perp}({\bm k},t=0) = {\hat k}_i \, {\hat k}_{\perp}^j\,\sigma_{ij}({\bm k},t=0) = \eta\,i\,k\,u_{\perp}({\bm k},t=0)\ .}
\label{eq:B.10}
\ee
This is not true at strictly $t=0$, as one can, in principle, prepare initial conditions for the velocity and the stress
tensor that are independent of each other. However, after a few collision times the relation will get established and
Eq.~(\ref{eq:B.10}) will hold as an effective initial condition with $t=0$ to be interpreted as $t=$ a few collision times.
The shear velocity is part of the hydrodynamic subspace, but the shear stress is not. To see how it enters the
hydrodynamic equations one must consider the underlying kinetic equation as an initial-value problem and project
onto the hydrodynamic subspace. This procedure was carried out in Ref.~\onlinecite{Kirkpatrick_Belitz_2023a}.
The result is that Eq.~(\ref{eq:4.17a}) becomes
\be
{u_{\perp}({\bm k},z) = M_{u_{\perp} u_{\perp}}({\bm k},z)\,(1 + \nu k^2\tau) u_{\perp}({\bm k},t^{(0)})}
\label{eq:B.11}
\ee
with $t^{(0)}$ on the order of a few collision times and $\tau$ a relaxation time associated with the shear viscosity
via $\eta \approx n\mu\tau$. Here we have ignored factors of $O(1)$ as well as a temperature dependence of the
non-hydrodynamic initial condition. Note that the latter is small compared to the hydrodynamic
one by a factor of $\nu k^2\tau \ll 1$, but of the same order in the gradient expansion as the viscous term in the
equation for the shear velocity.

\bigskip

\section{The heat diffusion coefficient}
\label{app:C}

Here we show how to obtain Eq.~(\ref{eq:2.35b}) from Eq.~(\ref{eq:2.34}). Consider fluctuations of the pressure,
\be
\delta p({\bm x},t) = \left(\frac{\partial p}{\partial T}\right)_{N,V} \delta T({\bm x},t) + \left(\frac{\partial p}{\partial\rho}\right)_{T,V} \delta\rho({\bm x},t)\ .
\label{eq:C.1}
\ee
If we ignore the fast pressure fluctuations, $\delta p = 0$, density fluctuations become proportional to temperature fluctuations, and if we use
the mass balance equation (\ref{eq:2.8}) we have
\bea
{\bm\nabla}\cdot{\bm u}({\bm x},t) &=& \frac{-1}{\rho} \bigl(\partial_t + {\bm u}({\bm x},t)\cdot{\bm\nabla}\bigr) \rho({\bm x},t) 
\nonumber\\
                                    &=& -\,\frac{(\partial p/\partial T)_{N,V}}{(\partial p/\partial\rho)_{T,V}}\,\bigl(\partial_t + {\bm u}({\bm x},t)\cdot{\bm\nabla}\bigr) T({\bm x},t)\ .
\nonumber\\                                    
\label{eq:C.2}
\eea
Using this in Eq.~(\ref{eq:2.34}) we find that the coefficient of the ${\bm\nabla}^2 T$ term becomes
\be
\kappa \left[c_V + \frac{T}{\rho}\left(\frac{\partial\rho}{\partial p}\right)_{T,V}\left(\frac{\partial p}{\partial T}\right)_{N,V}^2\right]^{-1}\ .
\label{eq:C.3}
\ee
Finally, if we use the thermodynamic identities (see Eqs.~(A20b) and (A27b) in Ref.~\onlinecite{Belitz_Kirkpatrick_2022})
\bse
\label{eqs:C.4}
\bea
\frac{1}{\rho}\left(\frac{\partial\rho}{\partial p}\right)_{T,V} &=& \frac{-1}{V}\left(\frac{\partial V}{\partial p}\right)_{T,N}\ ,
\label{eq:C.4a}\\
c_V - c_p &=& \frac{T}{V}\,\left(\frac{\partial V}{\partial p}\right)_{T,N} \left(\frac{\partial p}{\partial T}\right)_{V ,N}^2\ ,
\label{eq:C.4b}
\eea
\ese
we obtain Eq.~(\ref{eq:2.35b}).

\section{The collisionless regime as described by the Navier-Stokes equations}
\label{app:D}

The Chapman-Enskog derivation of the Navier-Stokes equations is valid only in the hydrodynamic regime,
$\vF k < 1/\tau$. However, the Navier-Stokes equations can, and historically have been, derived from much more
general arguments,\cite{Forster_1975, Ortiz_Sengers_2007} which suggests that they are more generally valid. 
Here we show how the results in the collisionless regime can be obtained from those in 
the hydrodynamic regime, at least in a scaling sense.

As one enters the collisionless regime from the hydrodynamic one, a crucial change is that the diffusive
modes, i.e., heat and shear diffusion, get replaced by pairs of propagating modes, see the discussion in
Ref.~\onlinecite{Belitz_Kirkpatrick_2022}. That is, a generic diffusion coefficient $D$ that can represent either
$\DT$ or $\nu$ effectively becomes a 
singular function of the wave number, $D \to \pm ic/k$, which turns a generic diffusion pole ${\cal D}$ into a propagating
mode,
\be
{\cal D}({\bm k},\omega) = \frac{1}{\omega + i D k^2} \to \frac{1}{\omega \mp ck}\ ,
\label{eq:D.1}
\ee
with $c \approx \vF$ the propagation speed. In what follows we perform a power-counting analysis, 
assuming that the diffusion coefficients scale as $D \sim \vF/k$. A much more complete analysis, especially
of the fluctuating force correlations, is needed to resolve issues regarding reality and signs.

Consider the commutator correlation $\chi_{TT}$ from {Eq.~(\ref{eq:3.6b})}, which has the structure
\bse
\label{eqs:D.2}
\be
\chi_{TT}({\bm k}) \sim \frac{T}{c_p} + \frac{\perpgradT^2}{\rho\, (D k^2)^2}\ .
\label{eq:D.2a}
\ee
Using $D \sim \vF/k$, dropping all constants of $O(1)$, and using the low-temperature result for the specific heat, we obtain
\be
\chi_{TT}({\bm k}) \to \frac{1}{\NF}\left[1 + \frac{\perpgradT^2}{\epsilonF^2 k^2}\right]\ ,
\label{eq:D.2b}
\ee
\ese
which is Eq.~(\ref{eq:3.13}). 

For analogous arguments concerning $\Ssym_{TT}$, consider the integral in Eq.~(\ref{eq:3.6b}):
\begin{widetext}
\bse
\label{eqs:D.3}
\be
\Ssym_{TT}(k)  =  \int_{-\infty}^{\infty} \frac{d\omega}{2\pi}\,\Ssym_{TT}({\bm k},\omega) = \Ssym_{TT,\text{eq}}({\bm k}) + \Ssym_{TT,\text{neq}}({\bm k})
\label{eq:D.3a}
\ee
where
\bea
\Ssym_{TT,\text{eq}}({\bm k}) &=& \frac{\DT k^2}{c_p}\,T \int_{0}^{\omega_0} \frac{d\omega}{\pi}\,\frac{\omega \coth(\omega/2T)}{\omega^2 + \DT^2 k^4}\ ,
\label{eq:D.3b}\\
\Ssym_{TT,\text{neq}}({\bm k}) &=& \perpgradT^2\,\frac{\nu k^2}{\rho} \int_{0}^{\infty} \frac{d\omega}{\pi}\,\frac{\omega \coth(\omega/2T)}{(\omega^2 + \DT^2 k^4)(\omega^2 + \nu^2 k^4)}\ .
\nonumber\\
\label{eq:D.3c}
\eea
\ese
\end{widetext}
The equilibrium contribution requires an interpretation: the integral diverges logarithmically in the ultraviolet and must be cut off.
The integrand is diffusive only up to frequencies $\omega \alt \DT k^2$, so as written the cutoff should be $\omega_0 \approx \DT k^2$.
At larger frequencies the integrand should be replaced by the free-electron propagator from Sec.~\ref{subsec:III.B} that leads to the
first term in Eq.~(\ref{eq:3.15a}). Combining these arguments, we see that in the long-wavelength limit, $\DT k^2 \ll T$, we recover
Eqs.~(\ref{eqs:3.14}). In the opposite limit, the two contributions mentioned above yield, apart from prefactors of $O(1)$,
\be
\Ssym_{TT,\text{eq}}({\bm k}) = \frac{T}{c_p}\,\DT k^2 + \frac{\vF k T}{c_V}\,\left[1 + O\left((\vF k\tau)^2\right)\right]\ .
\label{eq:D.4}
\ee
Upon using $\DT \sim 1/k$, both contributions are of the same order and we have
\be
S_{TT,\text{eq}}({\bm k}) \to \vF k/\NF\ ,
\label{eq:D.5}
\ee
in agreement with Eqs.~(\ref{eqs:3.15}). For the non-equilibrium part we find, in the same limit,
\be
\Ssym_{TT,\text{neq}}({\bm k}) = \frac{\perpgradT^2}{\pi\rho}\,\frac{\nu}{\nu^2 - \DT^2}\,\log(\nu/\DT)\,\frac{1}{k^2}\ ,
\label{eq:D.6}
\ee
and the scaling argument yields
\be
\Ssym_{TT,\text{neq}}({\bm k}) \to \frac{\vF k}{\NF}\, \frac{\perpgradT^2}{\epsilonF^2 k^2}\ ,
\label{eq:D.7}
\ee
again in agreement with Eqs.~(\ref{eqs:3.15}).

%\bibliography{Quantum_NESS}

\begin{thebibliography}{74}
\bibitem[{kpe()}]{kperp_footnote}
\bibinfo{note}{Note that ${\hat{\bm k}}_{\perp}$ does not change sign as ${\bm
  k} \to -{\bm k}$, see Eqs.~(\ref{eqs:A.2}).}
\bibitem[{BEC()}]{BEC_footnote}
\bibinfo{note}{In bosonic fluids at asymptotically low temperature
  Bose-Einstein condensation will lead to complications that require additional
  investigation.}
\bibitem[{Nav({\natexlab{a}})}]{Navier-Stokes_validity_footnote}
\bibinfo{note}{We note, however, that the Navier-Stokes equations remain valid
  in the collisionless regime in a scaling sense, see the remarks in
  Sec.~\ref{subsubsec:III.B.2} and the discussion in Appendix~\ref{app:D}.}
\bibitem[{car()}]{caret_footnote}
\bibinfo{note}{We use carets to denote operator-valued functions of space and
  time, and also to denote unit vectors.}
\bibitem[{sin()}]{single-particle_energy_footnote}
\bibinfo{note}{If the fermions are electrons in metals, then this form of
  $\epsilon_p$ implies a single parabolic band. This restriction is of no
  consequence for our purposes.}
\bibitem[{ext()}]{external_force_footnote}
\bibinfo{note}{In general the Uehling-Uhlenbeck equation also contains an
  external force that couples to ${\bm\nabla}_{\bm p} f_{\bm p}$. This is
  important, for instance, for studying the effects of a gravitational field on
  the fluid.}
\bibitem[{hba()}]{hbar_footnote}
\bibinfo{note}{In ordinary units, with $\hbar$ restored, the commutator
  correlation function must be defined as $\langle [\ ,\ ]_-\rangle/2\hbar$ to
  ensure that it exists in the classical limit, as we have done in
  Eqs.~(\ref{eq:1.3a}) and (\ref{eq:3.1b}). Alternatively, it can be defined in
  terms of Poisson brackets.}
\bibitem[{zer()}]{zero-sound_footnote}
\bibinfo{note}{In the collisionless regime the transverse velocity fluctuations
  are the propagating transverse zero-sound modes and hence not any softer than
  the longitudinal ones. We still neglect pressure fluctuations, for two
  reasons: (1) the coupling of the temperature gradient to the transverse
  velocity fluctuations is what yields the leading effects in the hydrodynamic
  regime and we are interested in the analogous effect in the collisionless
  regime. (2) Since all zero-sound modes have the same scaling behavior the
  coupling we neglect will lead to qualitatively the same effect as the one we
  keep. That is, the leading scaling behavior of the diverging correlation
  functions will be correct, but the prefactor will not. See
  Secs.~\ref{subsubsec:III.B.2} and \ref{subsubsec:V.B.2} for an additional
  discussion of this point.}
\bibitem[{loc()}]{local_equilibrium_footnote}
\bibinfo{note}{We stress again that, in real space, $T$, $\rho$, and $\nu$ are
  all position dependent due to the externally imposed temperature gradient in
  the NESS. The resulting effects are much smaller than the leading effect of
  the temperature gradient, which is the coupling of the shear velocity into
  the heat equation, and we neglect them by replacing these quantities with
  their spatially averaged values.}
\bibitem[{coe()}]{coefficients_footnote}
\bibinfo{note}{Note that the coefficients in Eqs.~(\ref{eq:A.6b}) and
  (\ref{eq:A.7}) can be rewritten to yield the expected derivatives, $Tn/c_p =
  (\partial T/\partial(s/n))_p$, $-(T/nc_p)(\partial n/\partial T)_{p,V} =
  (\partial T/\partial p)_{s/n}$, etc., as must be the case. To see this it is
  advantageous to rewrite the equations in terms of intrinsic quantities per
  particle: ${\tilde s} = s/n$, $\tilde v = 1/n$, and ${\tilde c}_{p} = c_{p}/n
  = T(\partial{\tilde s}/\partial T)_{p}$, ${\tilde c}_{V} = c_{V}/n =
  T(\partial{\tilde s}/\partial T)_{V}$.}
\bibitem[{Lap()}]{Laplace_trafo_footnote}
\bibinfo{note}{{Note that different authors define the Laplace transformation
  differently.}}
\end{thebibliography}

\end{document}